\documentclass[printer]{aa}

\usepackage{graphicx}
\usepackage{txfonts}
\usepackage{natbib}

\begin{document}
   \title{Hard X-ray spectral variability of the brightest Seyfert AGN in the Swift/BAT sample}

   \author{M.~D. Caballero-Garcia
          \inst{1,2}\fnmsep\thanks{E-mail address: mcaballe@physics.uoc.gr}
          \and
          I.~E. Papadakis \inst{3,1}
          \and
         F. Nicastro \inst{4,5,1}
          \and
          M. Ajello \inst{6}
          }

   \institute{Foundation for Research and Technology-Hellas, IESL, Voutes, 71110 Heraklion, Crete, Greece
         \and
 Physics Department, University of Crete, 710 03 Heraklion, Crete, Greece            
        \and
 University of Crete, Department of Physics and Institute of Theoretical \& Computational Physics, University of Crete, 710 03 Heraklion, Crete, Greece      
         \and
 Osservatorio Astronomico di Roma-INAF, Via di Frascati 33, 00040, Monte Porzio Catone, RM, Italy
         \and
              Harvard-Smithsonian Center for Astrophysics, 60 Garden Street, Cambridge, MA 02138, USA 
              \and
              SLAC National Laboratory and Kavli Institute for Particle Astrophysics and Cosmology, 2575 Sand Hill Road, Menlo Park, CA 94025, USA
              }


 
  \abstract
   {}
   {We used data from the 58 month  long, continuous  {\it Swift}/Burst Alert Telescope (BAT) observations of the five brightest Seyfert galaxies at hard X--rays, to study their flux and spectral variability in the 20--100 keV energy band. The column density in these objects is less than $10^{24}$ cm$^{-2}$, which implies that the  {\it Swift}/BAT data allow us to study the ``true" variability of the central source.} 
   {We used 2-day binned light curves in the (20--50) and (50-100) keV bands to estimate their fractional variability amplitude, and the same band 20-day binned light curves to compute hardness ratios and construct ``colour--flux" diagrams.  We also considered a thermal Comptonization model, together with a reflection component with constant flux, and produced model ``colour--flux" diagrams, assuming realistic variations of the model parameter values, which we then compared with the observed diagrams.}
   {All objects show significant variations, with an amplitude which is similar to the AGN variability amplitude at energies below 10 keV. We found evidence for an anti-correlation between variability amplitude and black hole mass. The light curves in both bands are well correlated, with no significant delays on time scales as short as 2 days. NGC~4151 and NGC~2110 do not show spectral variability, but we found a significant anti-correlation between hardness ratios and source flux in NGC~4388  (and NGC~4945, IC 4329, to a lesser extent). This ``softer when brighter" behaviour is similar to what has been observed at energies below 10 keV, and cannot be explained if the continuum varies only in flux; the intrinsic shape should also steepen with increasing flux.} 
   {The presence of significant flux variations indicate that the central source in these objects is intrinsically variable on time scales as short as $\sim 1-2$ days. The intrinsic slope of the continuum varies with the flux (at least in NGC~4388). The positive ``spectral slope--flux" correlation can be explained if  the temperature of the hot corona decreases with increasing flux. The lack of spectral variations in two objects, could be due to the fact that they may operate in a different ``state", as their accretion rate is less than 1\% of the Eddington limit (significantly smaller than the rate of the other three objects in the sample).}


   \maketitle

\section{Introduction}

Active Galactic Nuclei (AGN) are the most prominent and persistent X-ray sources in the sky. The current paradigm for the central source in these objects postulates the presence of a central black hole (BH) with a mass of $10^6-10^9$ M$_{\odot}$, and a geometrically thin, optically thick accretion disc that may extend to the innermost stable circular orbit around the black hole. Current unification schemes for radio-quiet AGN also postulate the presence of an obscuring molecular torus around the central source which prevents a direct view of the continuum and broad-line region in Seyfert 2 galaxies. The accretion disc  is thought to be  responsible for the broad, quasi-thermal emission component in the optical-UV spectrum of AGN (the so-called ``Big Blue Bump").  At energies above $\sim$2 keV, a power-law like component is observed in radio-quiet AGN. This is attributed to emission by a hot corona ($T \sim 10^8 - 10^9$ K) overlying the thin disc. The corona up-Comptonizes the disc soft photons to produce the hard ($E \sim 2 - 200$ keV) X-ray emission. X--rays from the corona can also illuminate the disc, producing the Fe ${\rm K}_{\alpha}$ line (6.4-6.97\,keV) and the ``reflection hump" at higher energies \citep{george91}. Reflection of the central radiation on the molecular torus often leads to an additional, possibly dominant, reflection component \citep{ghisellini94}. 

The AGN X--ray emission is strongly variable, even on time scales as short as a few hundred seconds, both in flux and in the shape of the observed energy spectrum. In most cases, radio-quiet objects show a positive spectral slope-flux relation, i.e. a Òsofter-when-brighterÓ behaviour, in the 2--10 keV band both on short and long time scales (see e.g. Sobolewska \& Papadakis 2009 and references therein). From a theoretical perspective, intrinsic slope variations are expected (Haardt, Maraschi \& Ghisellini 1997; Coppi 1999; Beloborodov 1999). For example, in the case of thermal Comptonization models, variations in the UV/soft X-ray photons can affect the slope of the X-ray spectrum and result in "softer" X--ray spectra when the source is bright. In fact such variations may have already been observed (see e.g. Nandra et al. 2000, Petrucci et al, 2000).

On the other hand, it has been argued that this apparent spectral steepening with increasing
source flux in the 2--10 keV band does not necessarily represent an intrinsic change in the spectral
slope. These variations can be explained if the power-law continuum is variable in flux, but not in shape, and the reflection component of the spectrum is constant. In this case, the superposition of this constant (in flux and shape) component and the continuum emission, of constant spectral slope and variable normalisation, can result in spectral softening when the flux increases (see e.g. Taylor, Uttley, \& McHardy 2003; Ponti et al. 2006;  Miniutti et al. 2007).Furthermore, the presence of an absorber whose either the column density, ionisation state and/or covering  factor of the source varies, can also explain the observed spectral variations in the 2-10 keV band of the radio-quiet AGN (see e.g. Turner \& Miller, 2009, for a review). Some cases of extreme absorption variations have indeed been reported so far in Seyfert galaxies, with  NGC~1365 being the most prominent case \cite{risaliti05,risaliti07,risaliti09}.

Clearly, the best option to avoid the effects of variable absorption to the shape of the X--ray spectrum is to study the ``hard" X--ray emission (i.e. X--rays at energies higher than $\sim 20$ keV). At these energies, only neutral absorbers with a column higher than $10^{24}$ cm$^{-2}$ can affect the shape of the intrinsic spectrum. Apart from these cases (which correspond to the so called ``Compton" thick sources) hard X-ray light curves  can in principle allow the study of the intrinsic continuum variability in AGN. AGN observations in hard X-rays have been performed the last twenty years by {\it CGRO},  {\it BeppoSAX} and {\it INTEGRAL}. The {\it Swift}/BAT all-sky survey in particular has provided us with continuously sampled light curves, which span periods as long as five years, for many AGN. 

Beckmann et al. (2007) have presented the results from a study of the first 9 month {\it Swift}/BAT light curves of 44 AGN. They found that $\sim 30\%$ of Seyferts exhibit significant hard X--ray variability on time scales of 20--150\,days, type 1 Seyferts are less variable than Seyferts 2, and a significant anti-correlation between luminosity and variability amplitude. More recently, Soldi~et al. (2010) reported the results from a preliminary study of the flux variability of 36 AGN using data from the first 5 years of {\it Swift}/BAT observations. Their results confirmed the hard X--ray variability -- luminosity anti-correlation detected by Beckmann et al. (2007) at high energies. They also showed that an anti-correlation between variability amplitude and BH mass may also exist for Seyfert galaxies. 

In this work, we present the results from a variability study of the five brightest radio-quiet AGN in the recently published catalogue of Baumgartner et al. (2011), using the 5 years long BAT light curves that the same authors provide in the 20--50 and 50--100 keV bands (the ``soft" and ``hard" bands, hereafter). Our main aim is to study the {\it spectral} variability of the sources with the use of ``hardness ratios", i.e. by simply dividing the hard over the soft band light curves. Such ratios have been extensively used in the past for the study of the AGN spectral variability in the 2--10/20 keV band.  Their biggest advantage is that they are entirely model--independent; if the hardness ratios are variable, then the spectral shape of the source {\it has} to be variable, irrespective of the underlying continuum spectrum, and of which spectral component is responsible of the observed variations. In addition, the presence (or absence) of a correlation between the hardness ratios and the source flux can indicate which model components vary (or not) in AGN. 

We restricted our study to the 5 brightest Seyferts in the current {\it Swift}/BAT catalogue because they are bright enough for an accurate estimation of their  soft and hard band fluxes on time scales as short as 20 days. As a result, we are able to use the hardness ratios to search for low-amplitude spectral variations on these time scales, almost continuously, over a period of 5 years. This would not be possible to achieve with the study of energy spectra extracted over periods as short as $\sim$ one month, due to low signal-to-noise ratio. In addition, the signal-to-noise ratio of even the 1--2 day binned soft band light curves of these objects is high enough to study their flux variations over a broad range of time scales, i.e. from years down to almost a day.

The sample and the light curves we used are described in Section \ref{sample}.  Our results from the flux and spectral variability analysis are reported in Sections \ref{fluxvar} and \ref{specvar}. We  discuss possible implications of our results in Section \ref{discuss}, and we present our conclusions in Section \ref{discuss}.

\section{The sample and data analysis}  \label{sample}

The Burst Alert Telescope (BAT; Barthelmy et~al.2005)  on-board the {\it Swift} \citep{gehrels04} is sensitive to X--ray photons in the 14-195\,keV energy range. Baumgartner et al. (2011) released a catalog of sources detected in the first 58 months of BAT observations. It consists of 1092 sources, detected at  a significance level of at least $4.8{\sigma}$.  The majority of the sources in this catalogue are AGN (with 519 objects classified as Seyferts). We chose to study  the 5 Seyferts which have the highest flux in  this {\it Swift BAT 58-Month Hard X-ray Survey} catalogue, among all radio-quiet AGN. Source names and their hard X-ray fluxes are listed in Table \ref{tsample}.  A summary of previous findings, regarding the hard X--ray emission of these objects, is presented in the Appendix.

\begin{table*}
\caption{The objects studied in this work. Average fluxes in the 14--195\,keV band  are taken from Baumgartner et~al. (2011). Columns 3
and 4 list the BH mass measurement we use in this work, and the references for these values. The fractional variability amplitudes, as well as $\chi^2$ values when we fit the light curves with a constant (and number of degrees of freedom), are listed for the 20--50\,keV band light curves (in parenthesis for the hard 50--100\,keV light curves). They are calculated using light curves binned in 2 days (1 day in the case of NGC~4151; see text for details).}         
\label{tsample}     
\centering          
\begin{tabular}{c c c c c c }   
\hline
Name(Type) & Flux     & ${\rm F}_{\rm var}$ &   ${\chi}^{2}/$(d.o.f)    &  BH$_{\rm mass}\,({\rm M}_{\odot})$     &   Ref.                 \\  
    &                        (${\times}10^{-10}$erg${\rm s}^{-1}{\rm cm}^{-2}$) &                    &                    &                 &                \\  
\hline                       
NGC~4151(S1.5)   &  $5.33$  &   $0.370{\pm}0.005$                 &   $11058/1326$                   &  $4.6^{+0.6}_{-0.5}{\times}10^{7}$  &   $1$                  \\    
    &                                               &  ($0.37{\pm}0.01$)         &   ($4272/1326$)   &                 &              \\
NGC~4945(S2)   &  $3.01$  &   $0.56{\pm}0.02$                   &   $2491/568$                       &  $1.4{\times}10^{6}$                &  $2$          \\
    &                                               &  ($0.43{\pm}0.03$)           &   ($1096/568$)  &                 &                 \\
NGC~2110(S2)   &  $2.97$  &   $0.43{\pm}0.02$                   &   $2823/585$                       &  $2{\times}10^{8}$                  &   $3$        \\
    &                                               &  ($0.45{\pm}0.03$)           &   ($1256/585$)  &                 &              \\
IC~4329(S1.2)   &  $2.90$  &   $0.35{\pm}0.02$                   &   $1403/526$                      &  $1.3^{+1.0}_{-0.3}{\times}10^{8}$  &   $4$              \\
    &                                               &  ($0.32{\pm}0.05$)           &   ($703/526$)   &                 &                  \\
NGC~4388(S2)   &  $2.76$  &   $0.47{\pm}0.01$                   &   $2985/658$                       &  $6{\times}10^{6}$                  &   $3$          \\
    &                                               &  ($0.48{\pm}0.02$)           &   ($1336/658$)  &                 &                 \\
\hline                        
\end{tabular}
\tablefoot{The references for the BH mass estimates are as follows: 1: Bentz et~al. (2006), 2: Greenhill et~al. (1997), 3: Woo \& Urry (2002), 4: Markowitz (2009)
}
\end{table*}

\begin{figure}
   \centering
    \includegraphics[bb=15 15 612 740,width=7.7cm,angle=270,clip]{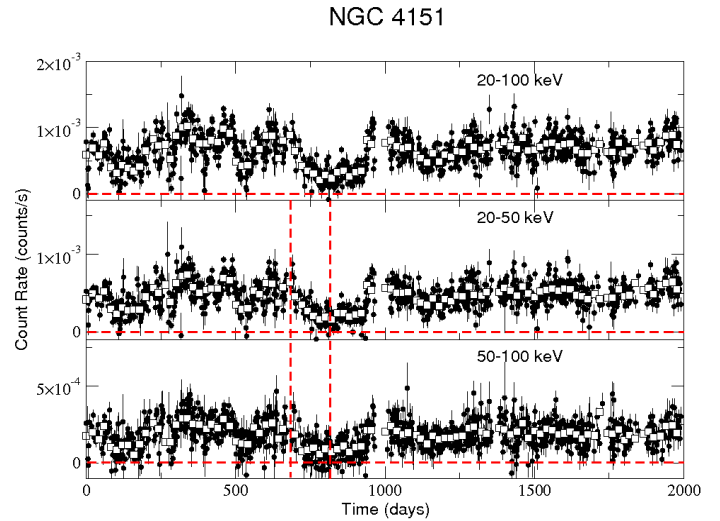}
  \caption{{\it Swift}/BAT light curves in the 50--100 keV (bottom panel),  20--50 keV (middle panel),  and full band (top panel) for the sources we studied.  Solid circles and open squares indicate the 2\,d (1\,d in the case of NGC~4151), and the 20\,d-rebinned light curves, respectively. The time axis counts days since the start of the {\it Swift}/BAT observations. The dashed vertical lines in the NGC~4151 plot indicate the case of a flux variation whose evolution in the hard band is faster than the respective flux change in the soft band (see section 3.1 for details).}
\label{light_curves}%
\end{figure}

\begin{figure}
   \centering
   \addtocounter{figure}{-1}
    \includegraphics[bb=15 15 612 740,width=7.7cm,angle=270,clip]{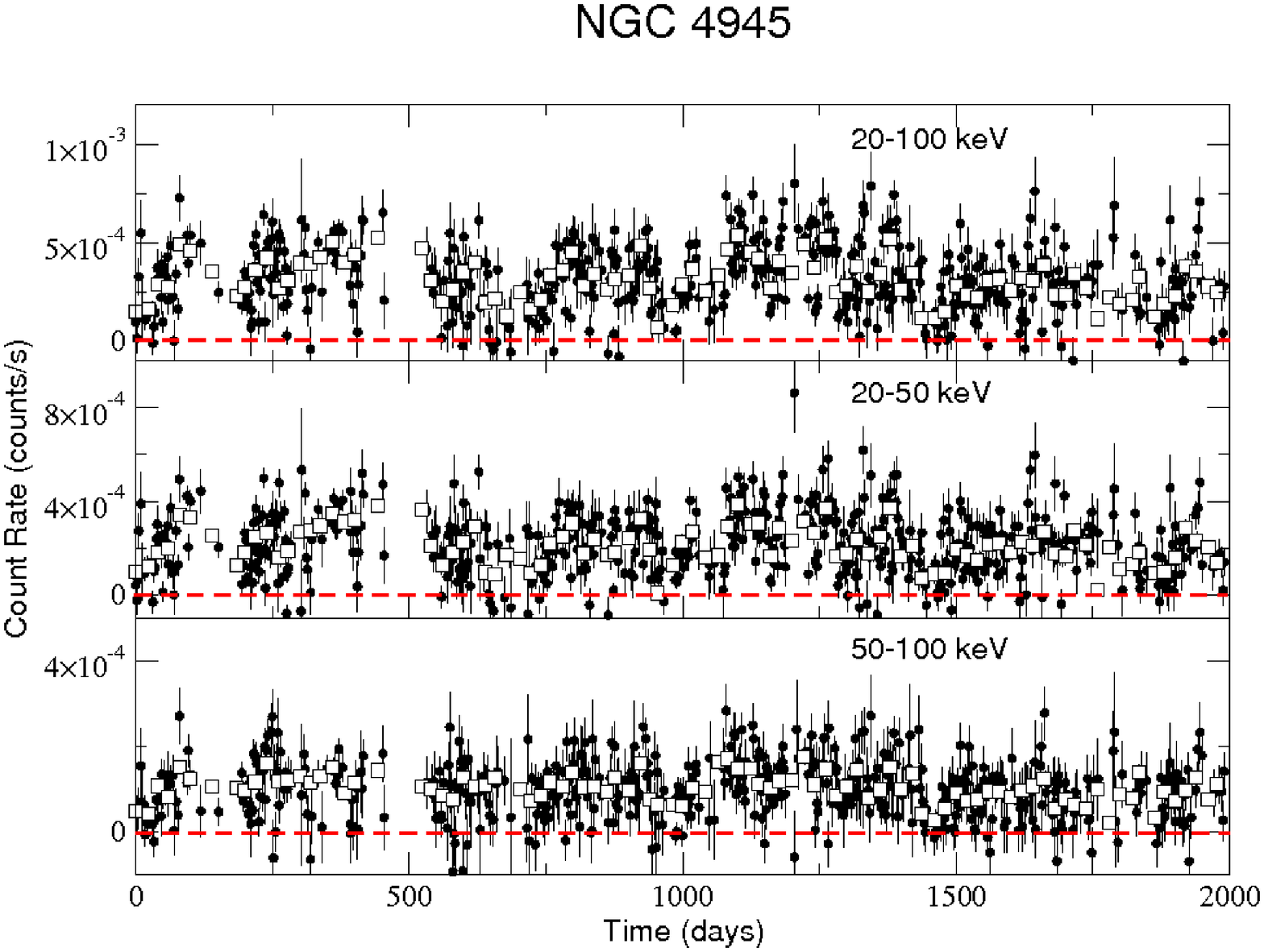}
   \caption{(Continued.)}
\end{figure}

\begin{figure}
   \centering
   \addtocounter{figure}{-1}
    \includegraphics[bb=15 15 612 740,width=7.7cm,angle=270,clip]{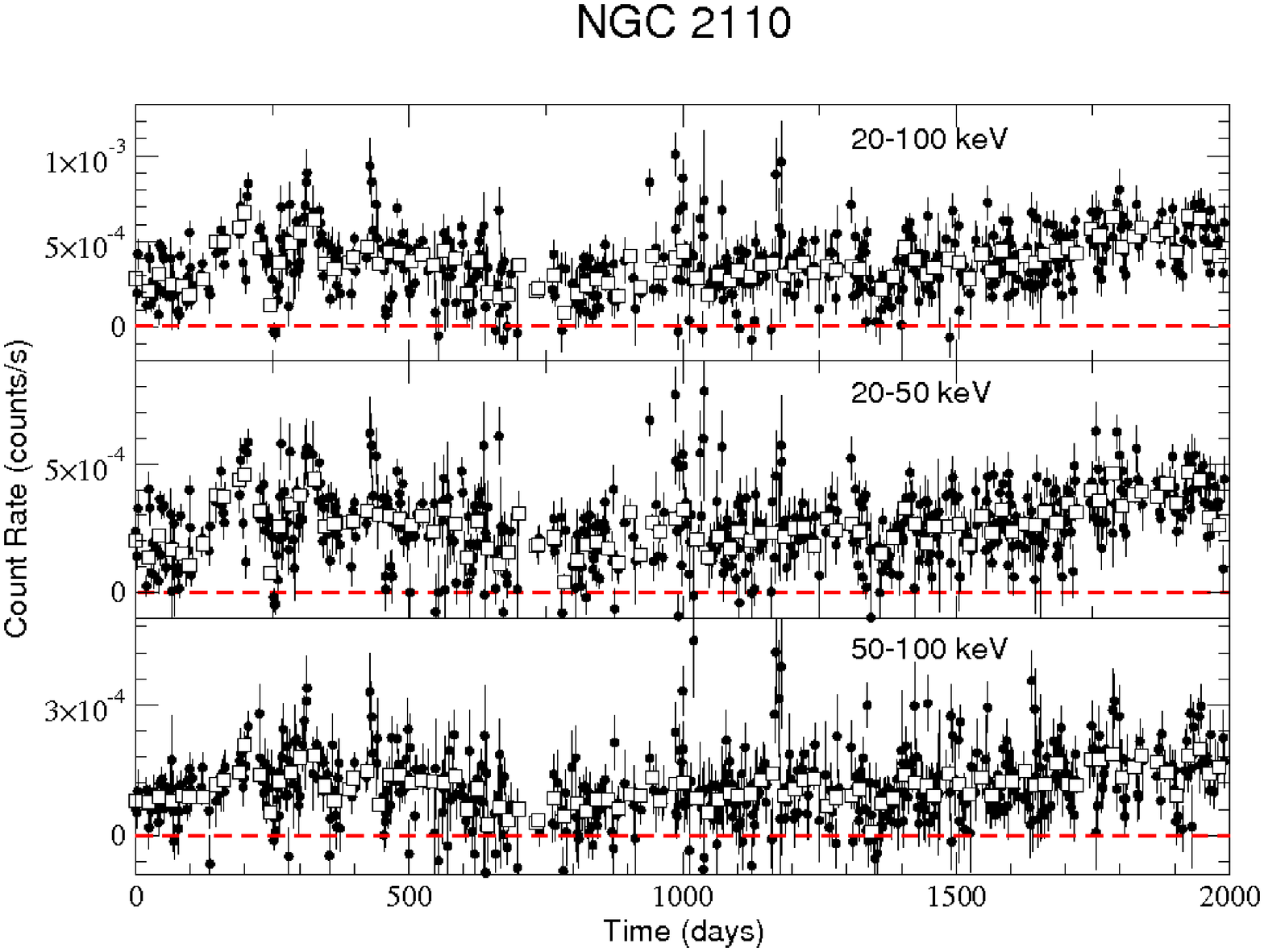}        
   \caption{(Continued.)}
\end{figure}

\begin{figure}
   \centering
   \addtocounter{figure}{-1}
    \includegraphics[bb=15 15 612 740,width=7.7cm,angle=270,clip]{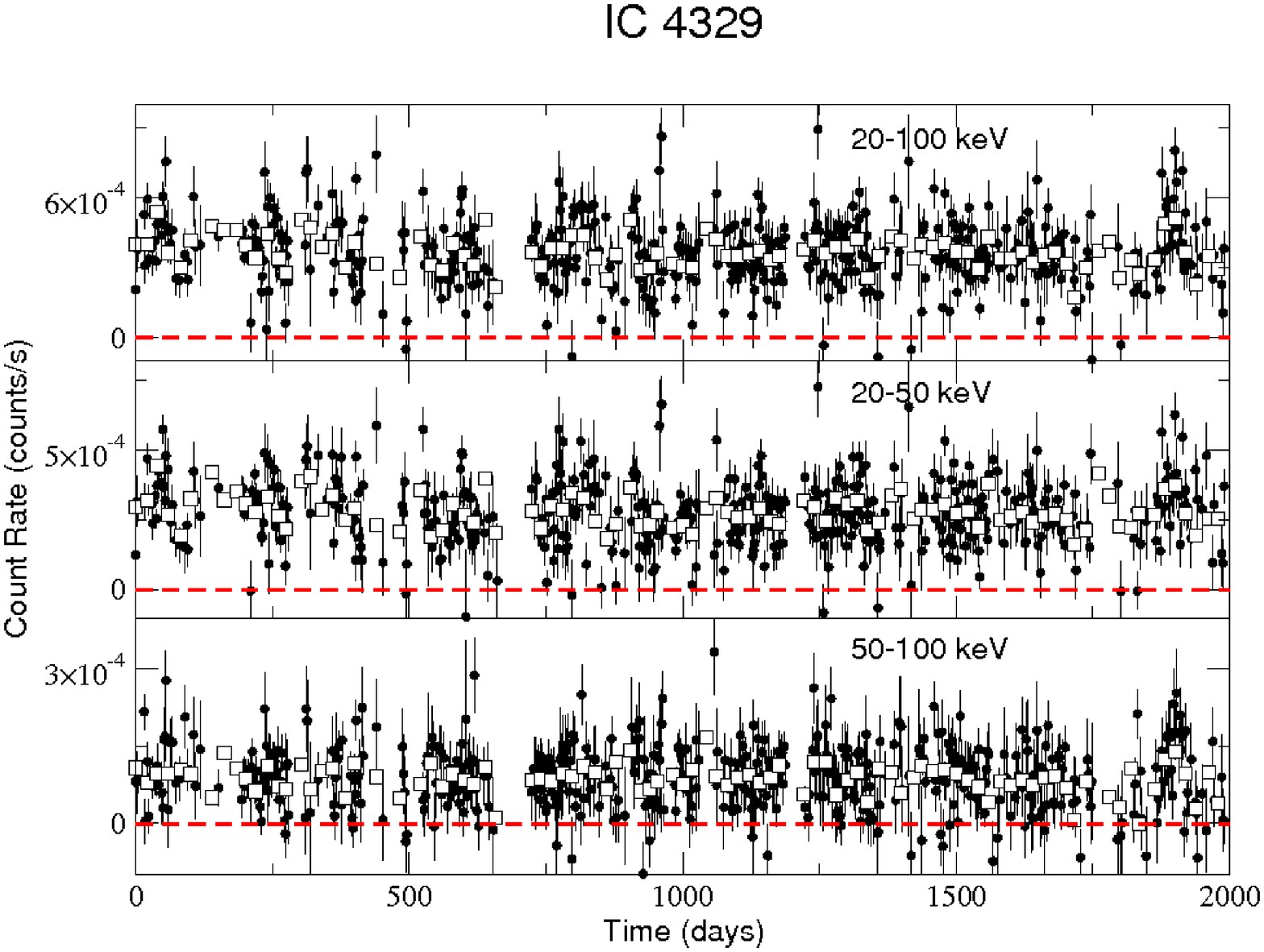}
   \caption{(Continued.)}
\end{figure}

\begin{figure}
   \centering
   \addtocounter{figure}{-1}
    \includegraphics[bb=15 15 612 740,width=7.7cm,angle=270,clip]{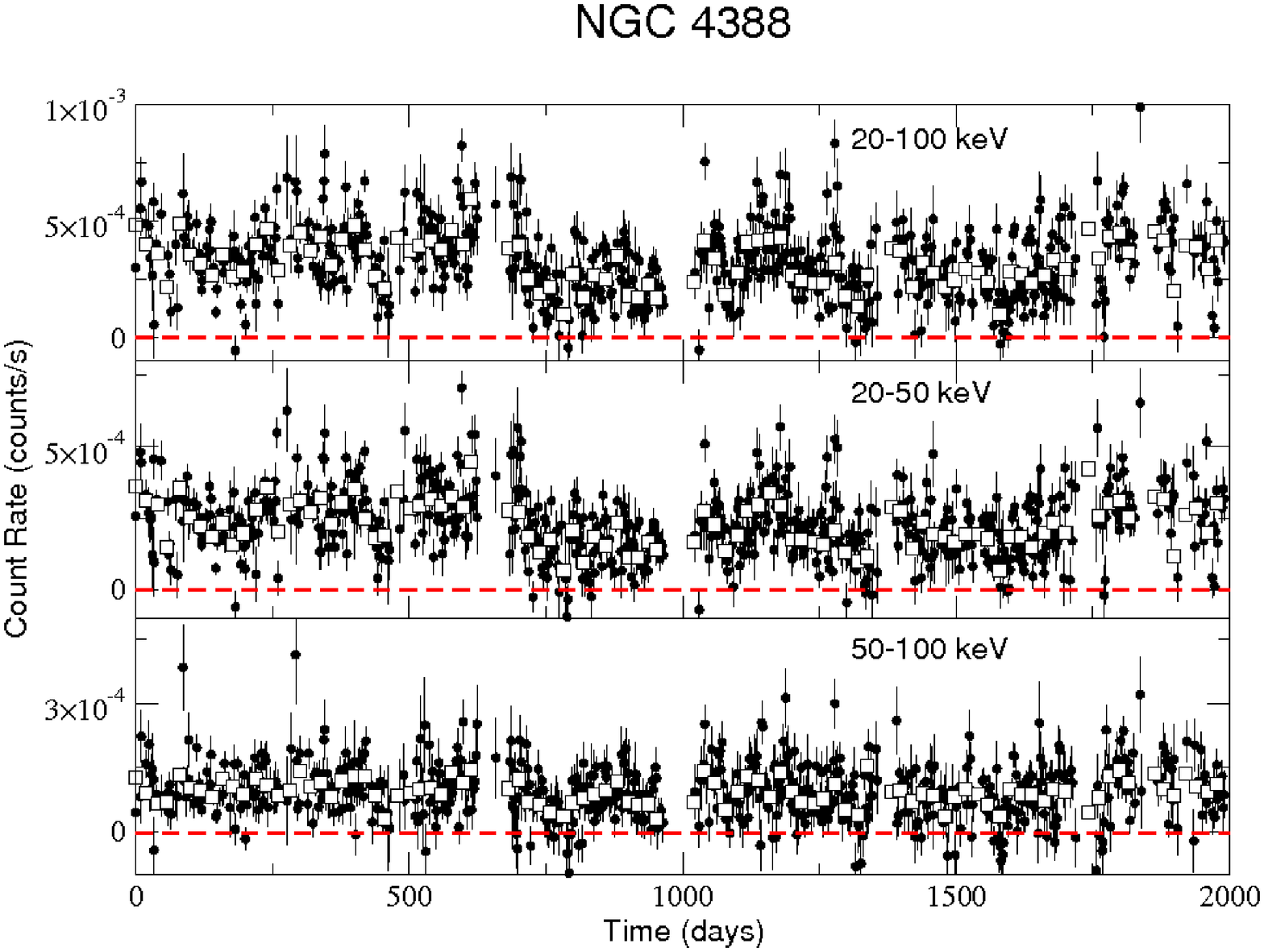}
   \caption{(Continued.)}
\end{figure}

Baumgartner et al. (2011) also provide light curves in eight energy bands: 14--20, 20--24, 24--35, 35--50, 50--75, 75--100, 100--150, and 150--195\,keV. These light curves are available from HEASARC\footnote{http://heasarc.gsfc.nasa.gov/docs/swift/results/bs58mon/}. They were extracted from the individual snapshot images from each ${\sim}5$\,min observation, and the reported count rates are corrected for off-axis effects. We added the count rate of the second, third and fourth band light curves to produce a combined light curve in the 20--50 keV energy band (the ``soft" band light curve hereafter; the error on the final count rate was calculated using the usual error propagation rules, i.e. Bevington 1969). In a similar way, we produced 50--100 keV (``hard" band, hereafter), and the ``full" band (i.e. 20--100 keV) light curves, by adding all the individual light curves in the respective energy ranges. 

The resulting light curves were then re-binned to 2 days and 20 days (in the case of NGC~4151, we used a bin size of 1 day, as this is the brightest source in our sample). We produced these light curves by estimating the weighted mean (and its error) of all the points within each bin, provided there were at least 20 points contributing to the estimation of the mean in it. The average ``count rate over error"  ratio (which is representative of the average ``signal-to-noise ratio" of the light curve) of the hard band light curves is $\sim 4$ for NGC~4151 and $\sim 2-2.5$ for the other objects. The ratio increases to $\sim 7$ and $\sim 4$ for the soft band light curves of NGC~4151 and the other objects, respectively. Given these values, the 2-d binned light curves at hand can 
reveal variations which have an amplitude at least $\sim 25$\% (50\%) of the mean flux of the sources in the soft (hard) band. 

The same average ratio increases to 15 and $\sim 5-7$ for the 20-d binned hard band light curve of NGC~4151 and of the other objects, respectively, and it is even higher ($8-27$) for the 20-d binned, soft band light curves. Given these values, the expected average signal-to-noise ratio for the hardness ratios is $\sim 11$ for NGC~4151 and larger than $\sim 4.5$ for all the other objects. These values imply that, even for the lowest signal-to-noise ratio light curves, we will be able to detect {\it spectral} variations which have an amplitude at least $\sim 20$\% of the mean hardness ratio.

\section{Flux variability} \label{fluxvar}

The soft, hard and total band light curves are plotted in Fig. \ref{light_curves}. They cover ${\sim}5$\,years of almost continuous observations of the objects, and as we argued in the previous section, their length, sampling pattern and relatively high signal-to-noise ratio make them ideal for the study of the hard X--ray flux and spectral variations of the sources. A visual inspection of the light curves indicates that all objects show significant variations. IC~4329 shows a gradual flux decay by a factor of $\sim 2$ in the first 2 years of the {\it Swift} monitoring (most pronounced in the total band light curve), and then its mean source flux level remains roughly constant with time. Other sources show larger amplitude variations on shorter time scales. For example, variations by a factor of $\sim 5$, in less than year, can be easily seen in the light curves of NGC~4151 and NGC~4388. 

In order to quantify the presence of significant variations in the light curves we performed a simple $\chi^2$ test when fitting the data to a constant. The results are listed in the last column of Table 1, together with the number of the degrees of freedom (dof). The reduced $\chi^2$ values imply highly significant variations. The smallest $\chi^2_{\rm red}$ value is 1.34 in the case of the hard band light curve of IC~4329, and, even in this case, the probability of the flux being constant is less than $4\times 10^{-4}$. To quantify the average variability 
amplitude of the sources we computed the fractional root mean square variability amplitude, F$_{\rm var}$, of the light curves. Following Vaughan et al (2003), this quantity is
defined as follows:


${\rm F}_{\rm var}=\sqrt{(S^2-\overline{\sigma}^2_{\rm err})/\overline{x}^2}$, 


\noindent where $S^2$ and $\overline{x}$ are the observed variance and mean of the light curve, and
$\overline{\sigma}^2_{\rm err}$ is the mean square error of all the measurement errors, $\sigma_{\rm err,i}$, of
each light curve point, i.e. ($\overline{\sigma}^2_{\rm err}=(\sum_{ i=1}^{N} \sigma^2_{\rm err,i})/N$. Since
the mean square error, $\overline{\sigma}^2_{\rm err}$, is representative of the contribution of the
measurement errors to the observed variance, $S^2$, F$_{\rm var}$ should be an estimate of the intrinsic
source variability amplitude over the time period sampled by the {\it Swift/}BAT light curves, i.e. $\sim 5$ years. 

\begin{figure}
   \centering
    \includegraphics[bb=65 65 612 752,width=7.5cm,angle=270,clip]{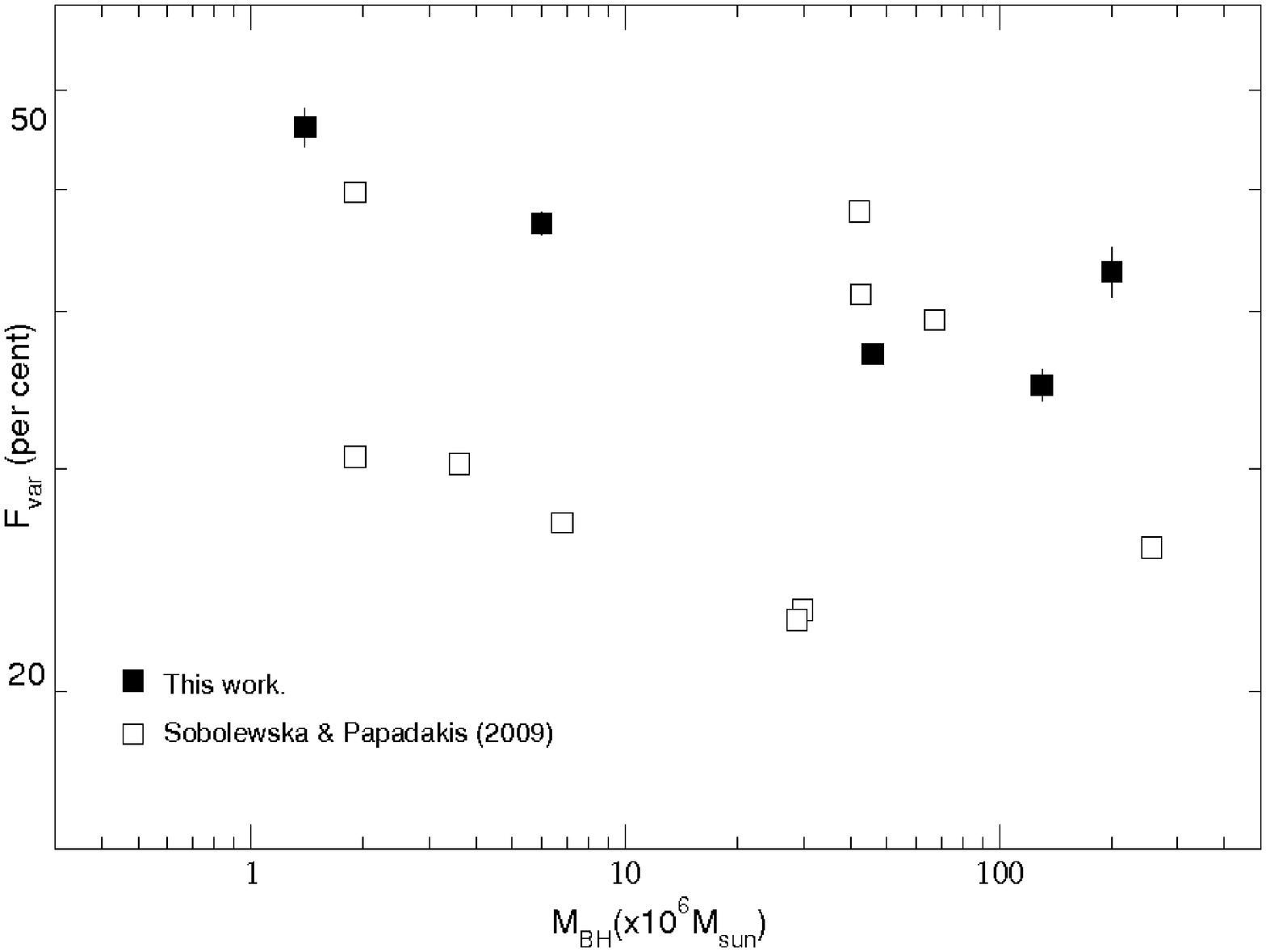}
  \caption{20--50 keV and 2--10 keV fractional variability amplitudes (black solid squares, this work, and open squares, from Sobolewska \& Papadakis, 2009, respectively) plotted as a function of BH mass.}
\label{rms}%
\end{figure}

The results are listed in the 3rd  column of Table~1. On average, the amplitude of the observed variations ranges from $\sim 30$\% up to $\sim 55$\% of the mean count rate. The average amplitude of all objects is comparable in both bands. This is not surprising, since the plots in Fig.\ref{light_curves} show that the same amplitude variations appear in the soft and hard band light curves of all objects.  Fig.~\ref{rms} shows a plot of the soft band $F_{\rm var}$ values vs BH mass for the objects in our sample (black, solid squares). The first thing to notice is that there appears an anti-correlation between variability amplitude and BH mass: smaller BH mass objects appear to be ``more" variable. The linear correlation coefficient is -0.83, but due to the small number of objects, the probability of an intrinsic correlation between $F_{\rm var}$ and BH mass is just 0.08. The same anti-correlation between variability amplitude and BH mass has also been observed using both long term (e.g. Papadakis 2004) and short term X--ray light curves (e.g. O'Neill et al, 2005; Zhou et al., 2010) in the 2--10 keV band.  

The average variability amplitude of the AGN we studied is comparable to the average variability amplitude of the AGN long term, 2--10 keV light curves. For example, Sobolewska \& Papadakis (2009) have recently studied the flux and variability properties of ten X--ray bright and radio-quiet AGN, using years long, 2--10 keV {\it RXTE} light curves.  The open squares in Fig.~\ref{rms} indicate their measurements.  Clearly,  at a given BH mass, the variability amplitude of the soft band {\it Swift} light curves is at least as large as that of the 2--10 keV light curves, despite the fact that the Sobolewska \& Papadakis (2009) {\it RXTE} light curves were longer than $\sim 7$ years, i.e. longer than the {\it Swift/}Bat light curves we use in this work.

   \begin{figure}
   \centering
   \includegraphics[bb=15 15 600 600,width=7.7cm,angle=270,clip]{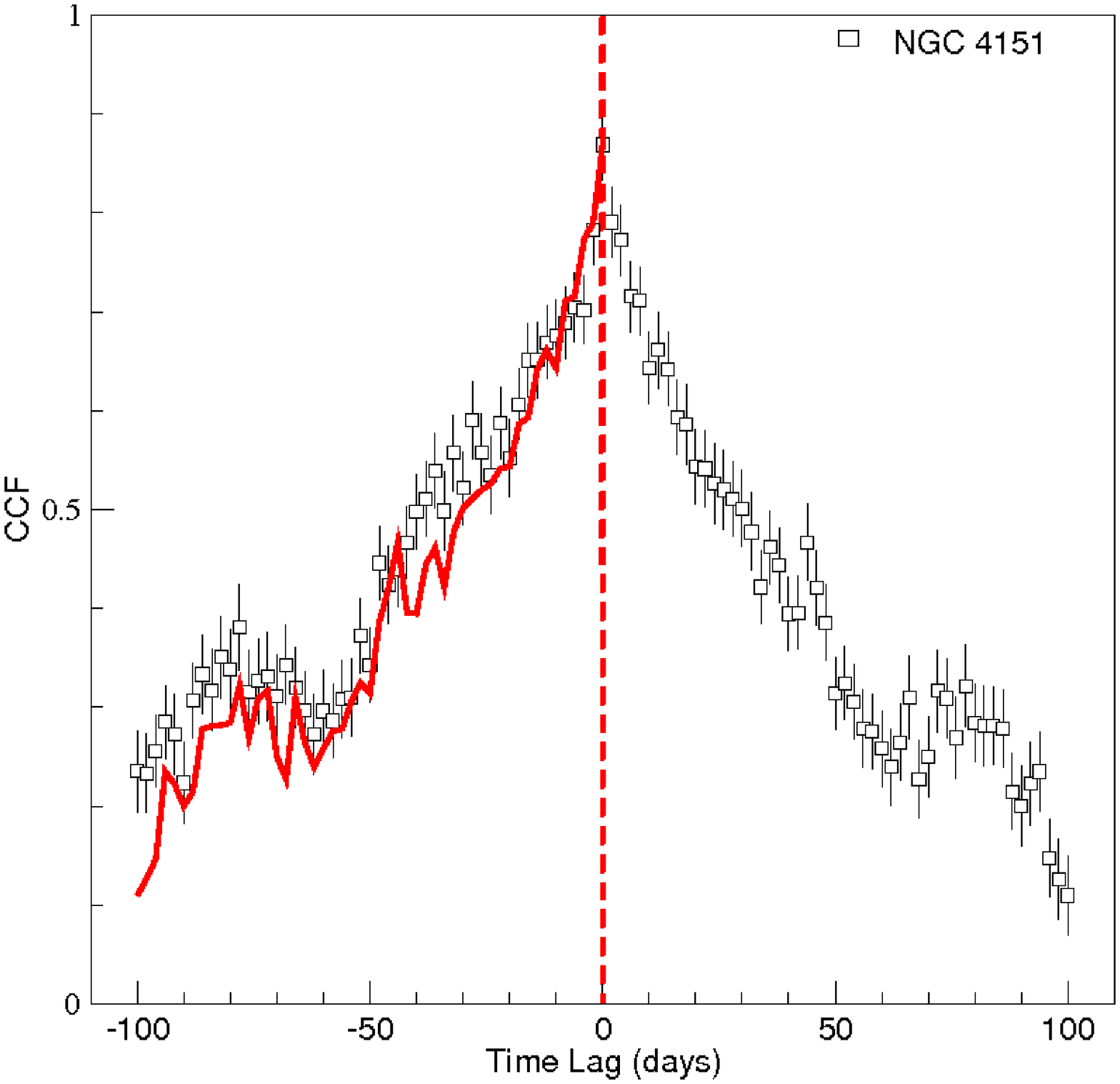}
   \caption{The NGC~4151 cross-correlation between soft and hard band light curves (open squares). The solid line (at lags $<0$) indicates the CCF values at positive lags (see text for details).}
              \label{4151ccf}%
    \end{figure}

\subsection{Cross-correlation analysis}
We calculated the sample Cross Correlation Function, CCF$(k)$, between the soft and hard band light curves following traditional methods (see e.g. Brinkmann et~al., 2003). The CCF was computed at lags $k=0,\pm1, \pm 2, \pm 3, ...$, days in the case of NGC~4151, while the lag bin size was equal to 2 days in the other objects. In all objects, the CCF peaks at zero lag with a maximum value which ranges between $\sim 0.5$ and $\sim 0.7$.  At larger lag values  the CCF decays to zero, with a similar rate in both the positive and negative lags. 

Given the excellent signal-to-noise of the NGC~4151 light curves, the resulting CCF for this object is much better defined over a large range of time lags. The NGC~4151 CCF is plotted in Fig.~\ref{4151ccf} (significant CCF peaks at positive lags in this plot means that the soft band variations are leading those in the hard band). 

The CCF peaks at a high value of $\sim 0.8$  at zero lag. The solid line at negative lags indicates the CCF($k>0$) values (i.e. it is a ``mirror line" of the CCF function at positive lags). A comparison between this line and the CCF values at negative lags suggests an asymmetry towards negative lags in the sense that the CCF$(k< -10$ days) values are larger than the CCF($k>10$ days) values. If real, this would suggest that, on time scales longer than $\sim 10$ days, the hard band {\it lead} the soft band variations. However, it is difficult to assess the significance of this result. 

We note though that there is at least one ``clear" case where the hard band variations do appear to ``lead" the variations in the soft band: the vertical dashed line in the middle and bottom panels of the NGC~4151 plot in Fig.~\ref{light_curves} indicates the start of a flux drop ``event" which appears simultaneously in both bands. However, although the hard band flux reaches its minimum flux level within $\sim 40-60$ days, the soft-band flux continues to decrease for at least 60-80 days longer. This faster flux decline in the hard band is an example of a hard band ``lead", which can explain the slight asymmetry towards negative lags we observe. 

\begin{table*}
\caption{Best-fit parameter values, $\chi^2$ values and number of degrees of freedom of Model A to the {\it HR}--time
curves (2nd and 3rd column), and of Model B \tablefootmark{a} to the Hardness-Flux plots (see text for details).
}
\label{tsample2}
\centering
\begin{tabular}{c c c c c c}
\hline
Name &  Model A     &                       & Model B   &             &      \\
     &  $\overline{HR}$   &  $\chi^2$/(d.o.f)      & $\beta$   & $\alpha$    & $\chi^2$/(d.o.f) \\

    &          &              &                        \\
\hline
NGC~4151  &  $0.40$      &  $215/97$     & $0.005{\pm}0.012$      & $0.391{\pm}0.003$   &  $212/96$                      \\
NGC~4945  &  $0.52$      &  $188/91$     & $-0.07{\pm}0.03$       & $0.461{\pm}0.011$   &   $124/90$                     \\
NGC~2110  &  $0.44$      &  $115/97$     & $-0.014{\pm}0.023$     & $0.419{\pm}0.008$   &   $106/96$        \\
IC~4329   &  $0.31$      &  $129/90$     & $-0.09{\pm}0.05$       &  $0.303{\pm}0.007$  &  $121/89$                      \\
NGC~4388  &  $0.43$      &  $174/92$     & $-0.13{\pm}0.03$       & $0.418{\pm}0.008$   &   $136/91$                     \\
\hline
\end{tabular}
\tablefoot{ (a) ${\rm HR}={\alpha}+{\beta}{\times}$ (20--100 keV Flux)}
\end{table*}

   \begin{figure*}
   \centering
   \includegraphics[bb=0 0 612 792,width=7.0cm,angle=270,clip]{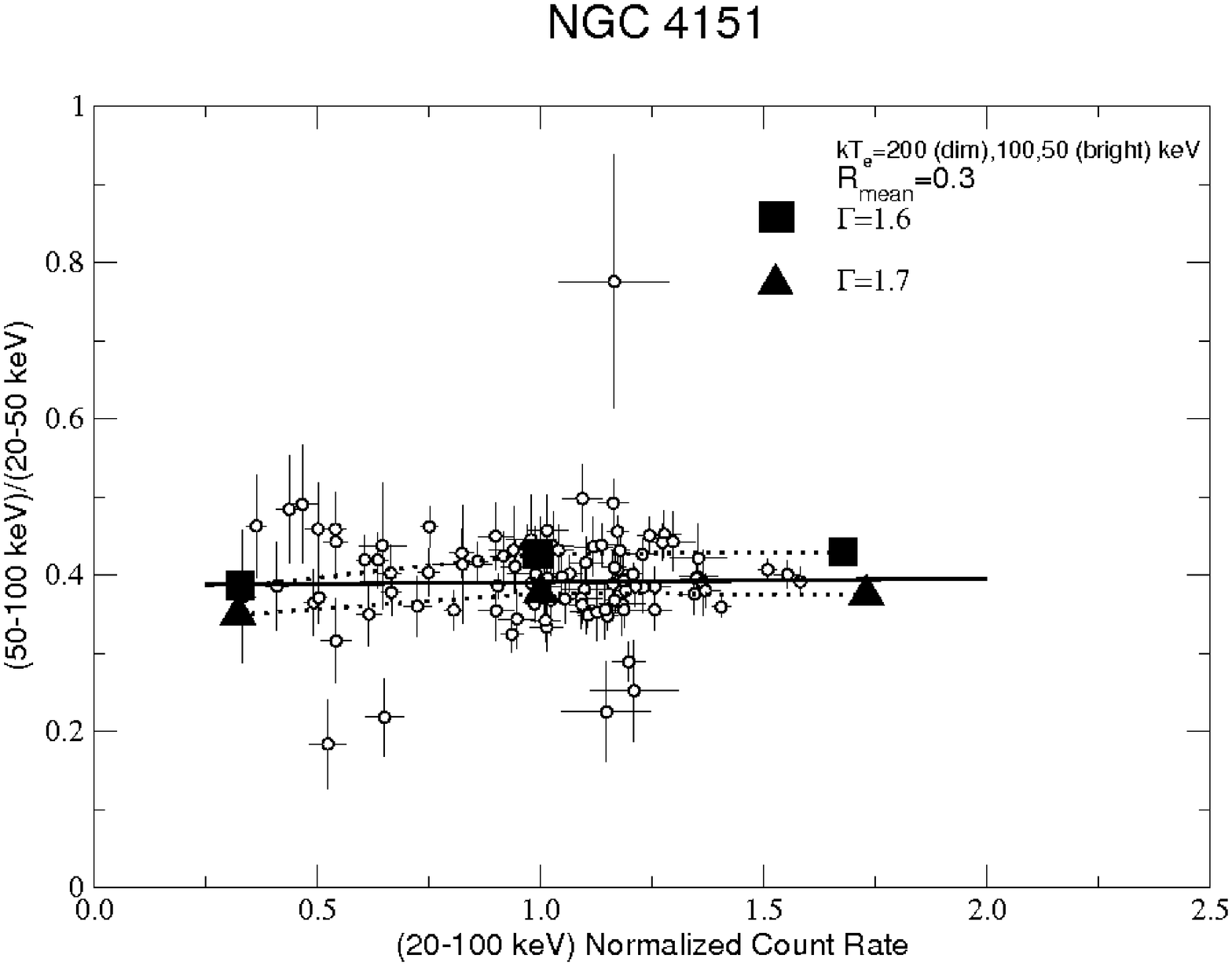}
   \includegraphics[bb=0 0 612 792,width=7.0cm,angle=270,clip]{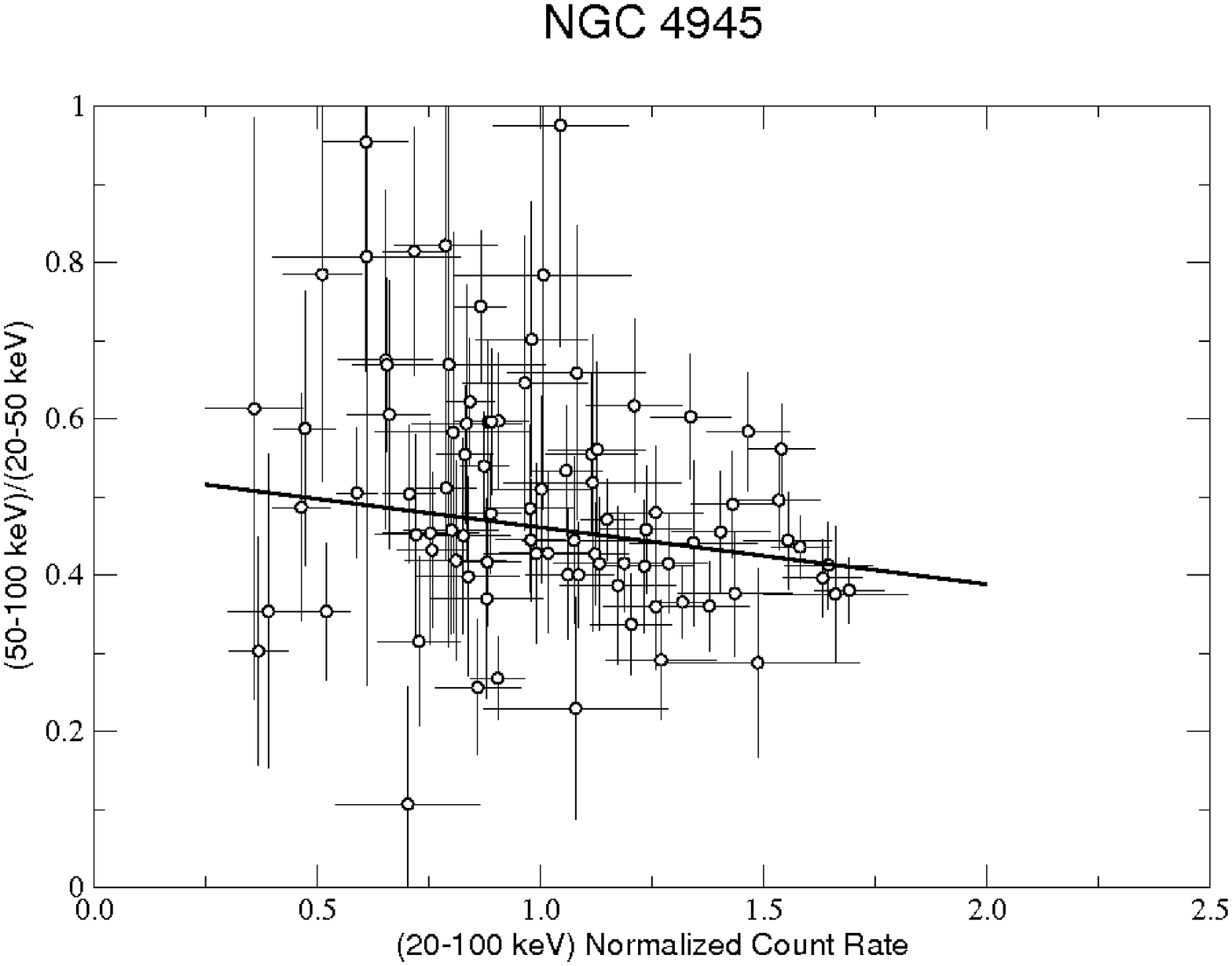}
   \includegraphics[bb=0 0 612 792,width=7.0cm,angle=270,clip]{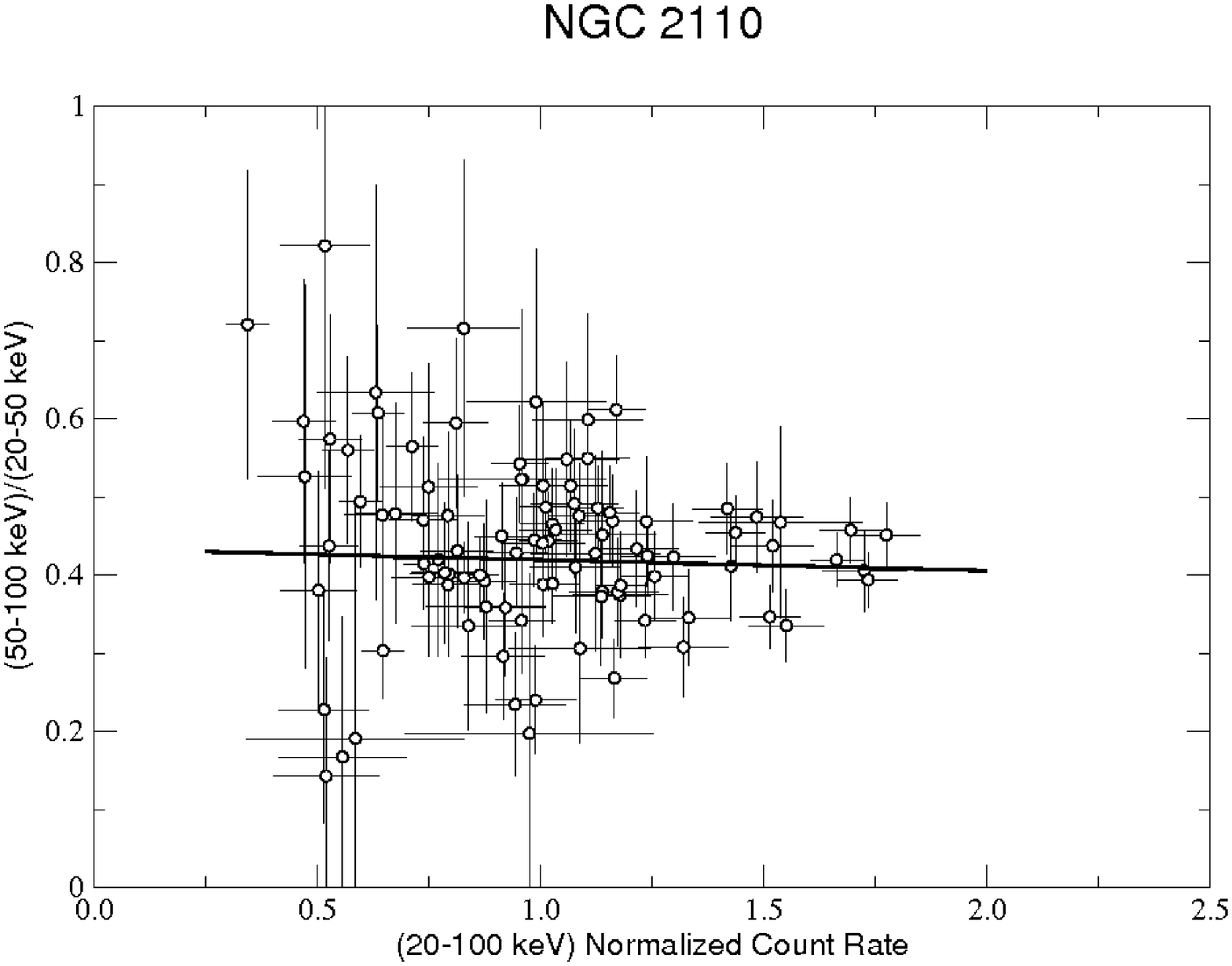}
   \includegraphics[bb=0 0 612 792,width=7.0cm,angle=270,clip]{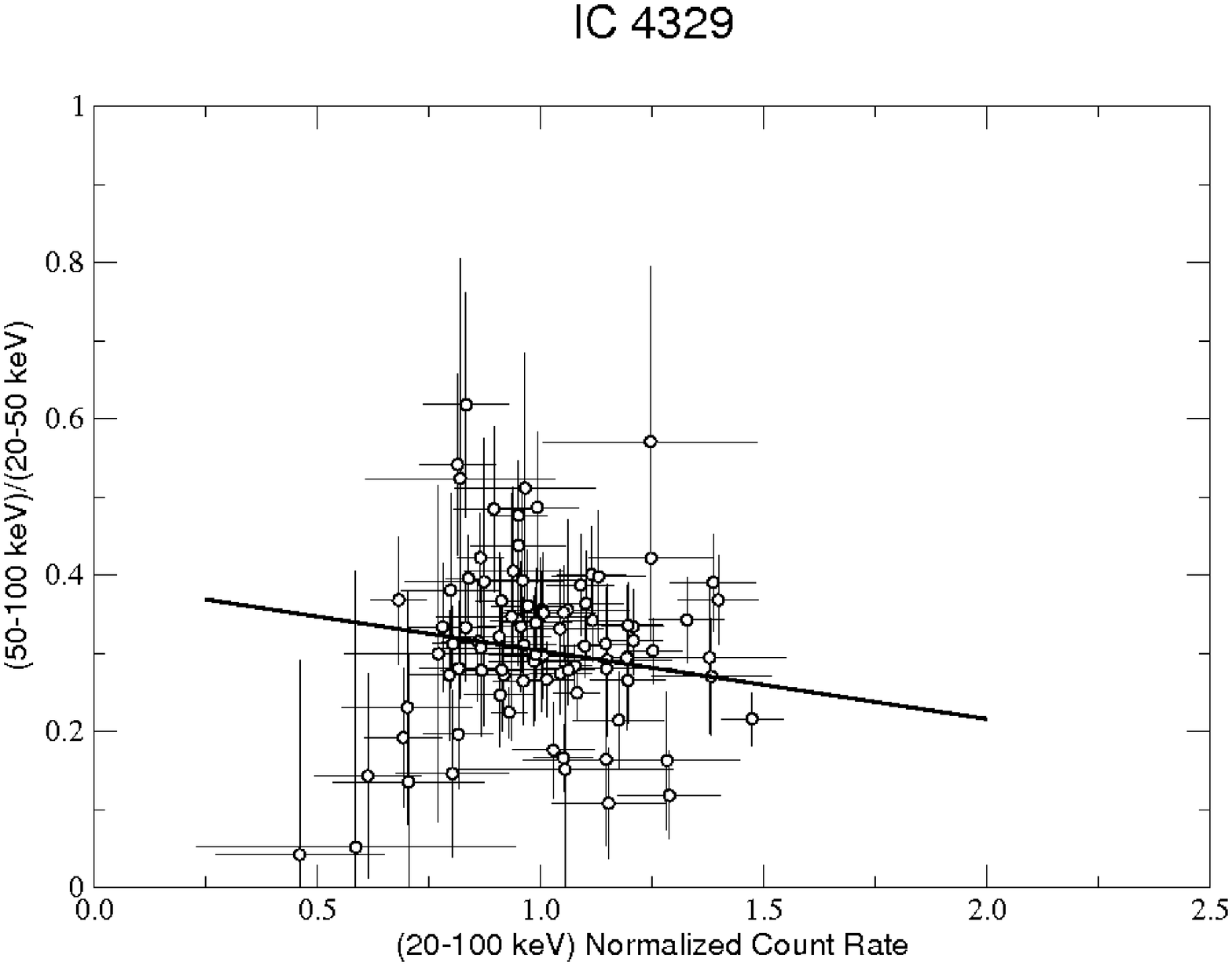}
   \includegraphics[bb=0 0 612 792,width=7.0cm,angle=270,clip]{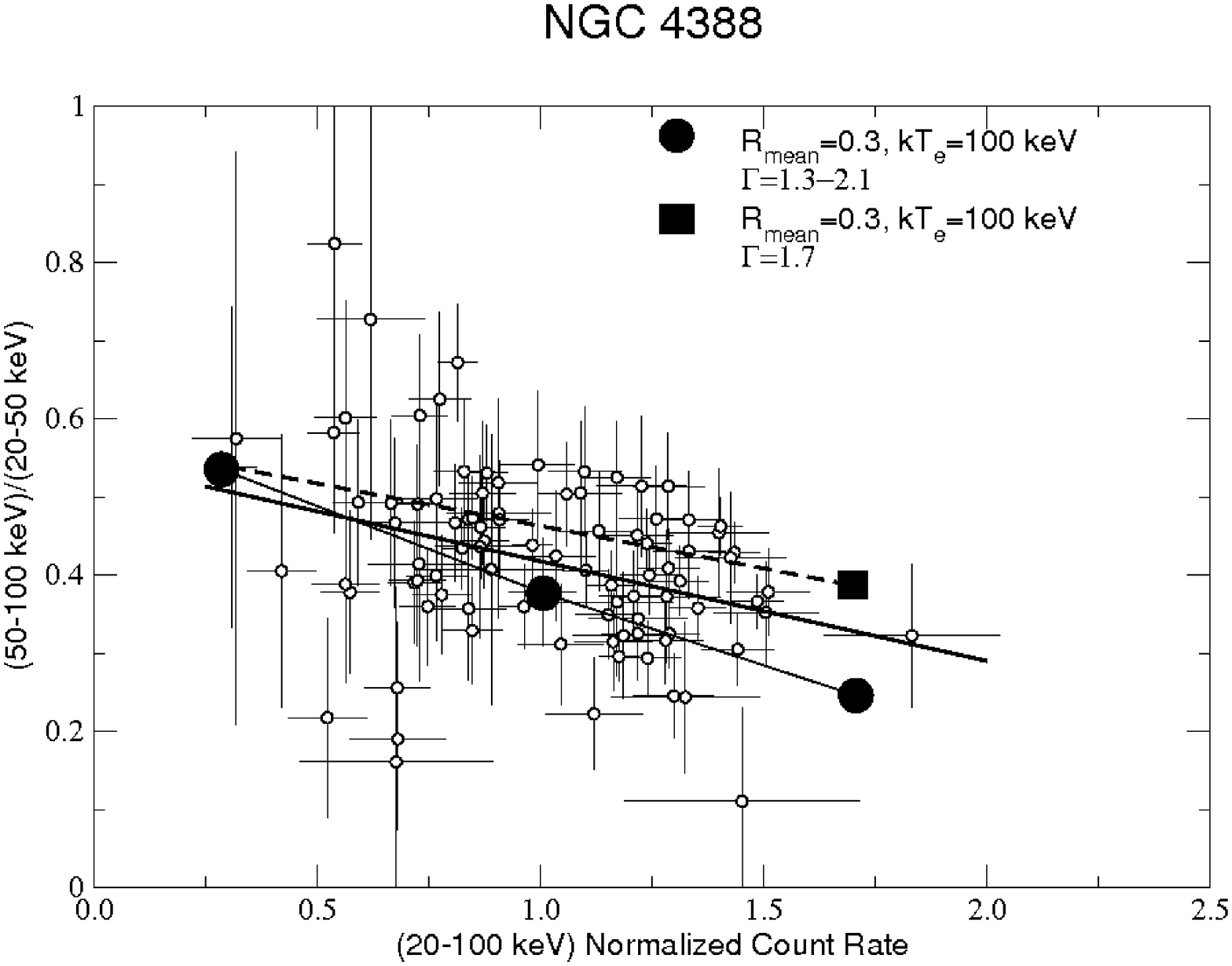}
   \caption{Hardness ratio (50--100\,keV)/(20--50\,keV) values, estimated using the 20\,d-binned light curves, plotted as a function of the total (20--100)\,keV normalized count rate. The solid lines show the best-fitting linear function to the data.}
    \label{colourflux}%
    \end{figure*}

\subsection{Hardness ratio analysis}

We used the 20\,d-rebinned soft and hard band BAT light curves ($CR_{\rm 20-50\,keV}$ and $CR_{\rm 50-100\,keV}$, respectively) to compute the hardness ratio, $HR=CR_{\rm 50-100\,keV}/CR_{\rm 20-50\,keV}$, and study the spectral variability of the sources. We assume below that: a) a model fits the data well if the null hypothesis probability is larger than $1$\%, and b) a model significantly improves the goodness of fit if the null hypothesis probability (i.e. ``both models fit the data equally well") is less than $1$\%.

First, we fitted the {\it HR}$-$time curves with a constant (hereafter ``model A"). This constant is
basically equal to the mean {\it HR} value, $\overline{HR}$. The resulting best-fit $\overline{HR}$ and
$\chi^2$ values, together with the number of degrees of freedom (dof), are listed in the second and third
column of Table 2. These values indicate that the probability of a constant $HR$ is less than 1\% in all 
objects, except for NGC~2110 where the null hypothesis probability is 10.1\%. 

We then produced ``colour--flux" diagrams, i.e. diagrams of the $HR$ values plotted as a function of the total band flux ($CR_{\rm 20-100\,keV}$). These diagrams are shown in Fig. \ref{colourflux} (the full band count rates in these plots are normalized to the respective mean count rate). The NGC~4151 colour--flux diagram shows that the $HR$ values cluster around $\sim 0.4$. The high $\chi^2$ value when we fit a constant to the NGC~4151 diagram is due to the presence of a few points, with small error bars, which scatter randomly around the mean $HR$. Although these points may indicate significant, intrinsic spectral variations which last for a short period of time (i.e. less than 20 days), the overall picture that emerges from the NGC~4151 diagram in Fig. \ref{colourflux} is that, despite the significant, large amplitude flux variations we observe in this object, the shape of the spectrum should remain roughly constant in this source. This is in agreement with the results form the recent work of Lubi{\'n}ski et~al. (2010).

On the other hand, the NGC~4388 colour--flux diagram (and perhaps the diagrams of NGC~4945 and IC~4329 as well) suggest flux related spectral variations: as the source flux increases, the $HR$ values appear to decrease. This is identical to what is observed in the X--ray bright, radio-quite Seyfert galaxies in the 2--10 keV band: the spectrum becomes ``softer" with increasing source flux.  In order to quantify the significance of this apparent trend, we fitted all colour--flux diagrams with a linear function of the form: ${\rm HR}={\alpha}+{\beta}{\times}CR_{\rm 20-100\,keV}$ (hereafter ``model B"; during the fit we took into account the errors on both variables, following Press et al., 1994). The model B best fit results are listed in the last columns of Table 2, and the solid lines in Fig. \ref{colourflux} indicate the best-fit lines.  

In the case of NGC~2110, model B appears to improve the goodness of fit to the $HR-$flux data of this source ($\Delta\chi^2=9$ for 1 dof; null hypothesis probability$=0.5$\%). However, as we mentioned above, a constant $HR$ model already provides an acceptable fit to the colour--flux diagram of this source. In addition, the model B best-fit slope is consistent with zero within the errors. We therefore conclude that, just like NGC~4151, we do not find significant evidence for spectral slope variations in NGC~2110. 

Contrary to this, the best-fit slope value of NGC~4388 is different than zero (the case of constant $HR$) at the 3.3$\sigma$ level. In addition, according to the $F-$test model B provides a significantly better fit to the data ($\Delta\chi^2=38$ for 1 dof; null hypothesis probability$=2.3\times10^{-4}$\%) when compared to model A. We therefore conclude that NGC~4388 shows significant spectral variations, with a ``softer when brighter" behaviour.

The model B best-fit slope of NGC~4945 and IC~4329 is negative, and consistent with the best-fit slope of NGC~4388 (within the errors). However, the negative slope in these objects is significant at just the $2.2\sigma$ and $1.8\sigma$ level. On the other hand, model B fits relatively well the colour--flux diagrams of these sources (null hypothesis probability$=1.1$\% and 1.4\%, respectively), while model A does not. In addition, the improvement of the model B best-fit, when compared to model A best fit, is highly significant in the case of NGC~4945 ($\Delta\chi^2=64$ for 1 dof; null hypothesis probability$=1\times10^{-7}$\%). This is not the case in IC~4329. The colour--flux diagram of this source is broadly similar to the diagrams of NGC~4945 and NGC~4388, but there are quite a few points which cluster around a value of $HR\sim 0.2$, irrespective of their flux. We conclude that there are indications for flux related spectral variations in NGC~4945 (mainly) and IC~4329, but they are not as significant as in NGC~4388. 

In summary, NGC~4151 and NGC 2110 do not show significant spectral variations, while we detect significant spectral variability in NGC~4388. The situation is less clear in NGC~4945 and IC~4329: their colour--flux diagrams are similar to NGC~4388, but the presence of spectral variations is not highly significant. 

\begin{figure}
   \centering
   \includegraphics[bb=33 49 717 528,width=9cm,angle=270,clip]{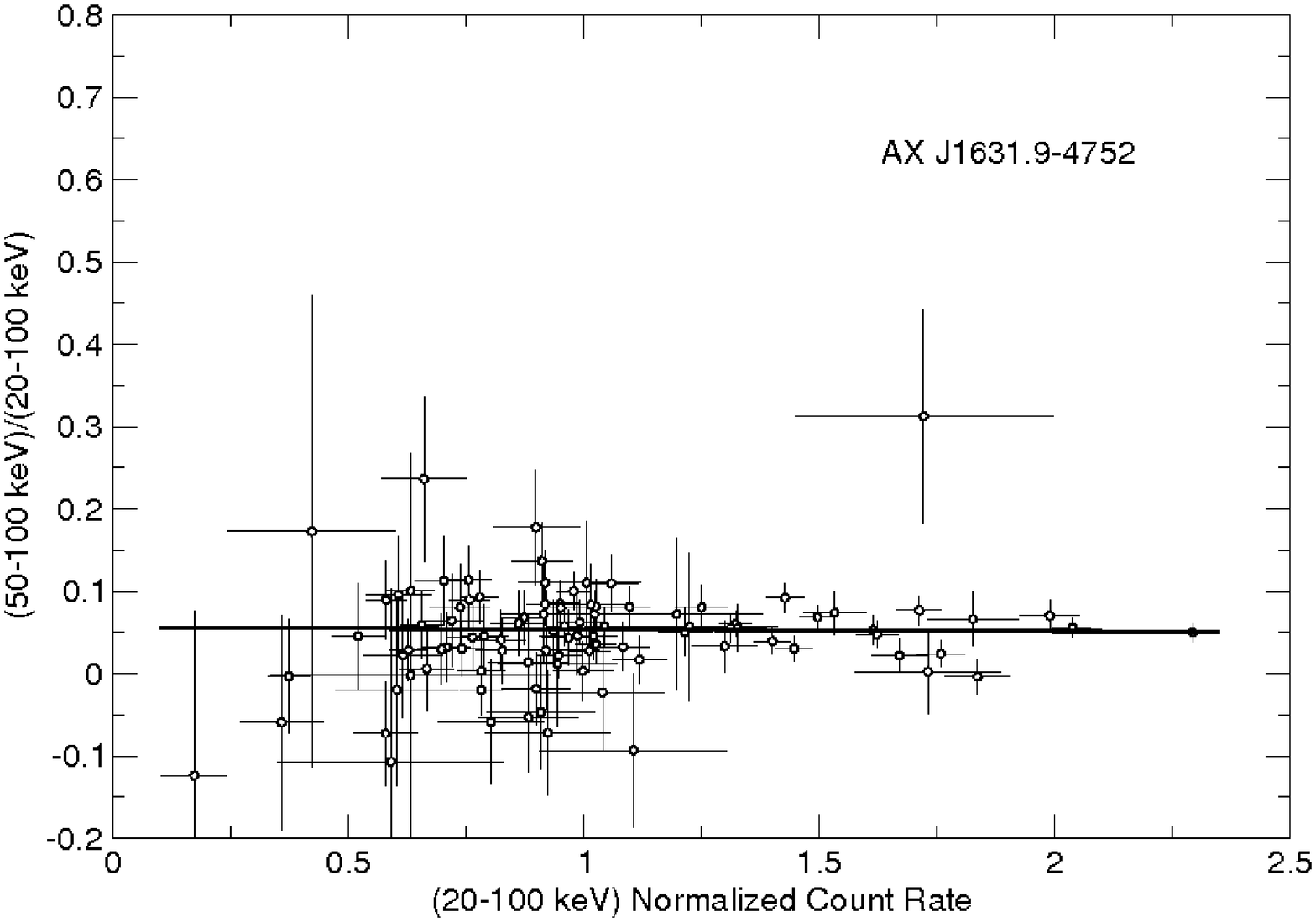}
   \caption{Colour--flux diagram of AX J1631.9-4752.}
              \label{ax}%
    \end{figure}

Since NGC~4388 is the lowest flux object in our sample, we investigated the possibility that the spectral variations we detect in this object may be due to unaccounted systematic uncertainties in its {\it Swift}/Bat light curves. For this purpose, we downloaded from HEASARC the {\it Swift/}BAT light curves of AX J1631.9-4752, a source with a 20--100 keV flux similar to the NGC~4388 flux. This is an X-ray pulsar in a High Mass X--ray binary. Its spectrum is well fitted with an exponentially cut-off power-law model, with a cut-off energy of $\sim 10$ keV. It remains constant (in shape) both during flaring and non-flaring periods (Rodriguez et al., 2006). Fig.~\ref{ax} shows the colour--flux diagram for this source. A few $HR$ values in this diagram are negative, because the source's flux is very low above 50 keV, due to the fact that its X--ray spectrum is very steep above above 20 keV. For the same reason, the $HR$ values are on average smaller than the $HR$ values of NGC~4388 and the other objects in our sample. The points along the $x-$axis indicate significant flux variations, whose amplitude is even larger than the amplitude of the X--ray variations we observe in NGC~4388. 

Despite these large amplitude variations, the colour--flux diagram of this source appears to be remarkably flat. The solid line in the same figure indicate the model B best-fit line to the data. The best-fit slope is -0.0023$\pm 0.007$, entirely consistent with the value of zero. We therefore conclude that the flux related spectral variations we detect in NGC~4388 are most probably intrinsic to the source, and are not caused by any systematic uncertainties. In fact, if the error on the model B best-fit slope is less than 0.01 for an object with no intrinsic spectral variations, and a count rate similar to (or less than) the count rate of the sources in our sample (like AX J1631.9-4752 for example), then the model B best-fit slopes in the case of the NGC~4945 and IC~4329 colour--flux diagrams should be indicative of significant spectral variations in these sources as well. 

\section{Spectral variability}   \label{specvar}

\subsection{Interpretation of the  ``colour--flux" diagrams}  \label{spe_flux}

Although the hardness ratios can be used to detect spectral variations in a model independent way, they cannot identify, unambiguously, which are the model components that are actually responsible for the observed variations. The main reason is that in most cases (including our work) the hardness ratios are computed using light curves over energy bands which are quite broad, and as a result multiple components contribute to the observed count rate. In order to investigate in a quantitative way the constrains that the colour--flux diagrams in Fig.~\ref{colourflux} can impose on current theoretical models, we performed the ``experiment" we describe below.

Thermal Comptonization is the most commonly accepted mechanism for the X--ray emission from radio-quiet AGN. We therefore considered a  ``Comptonization plus reflection" model, similar to what has been used numerous times in the past to fit the high energy X--ray spectra of many AGN, and we computed the expected hardness ratios when the model parameters were allowed to vary. We produced ``theoretical" colour--flux diagrams, which we then compared with the observed diagrams. 

We first used the thermally comptonized continuum model {\tt Nthcomp} in XSPEC \cite{zdziarski96,zycki99}. The main model parameters are the spectral slope, $\Gamma$, and the electron temperature, $kT_{e}$, which determines the high energy rollover of the spectrum. As for the other model parameters, we assumed a seed photon temperature of 10 eV, and a ``diskblackbody" input spectrum (none of these two parameters affect significantly the shape of the spectrum at energies $> 20$ keV). We then added a reflection component using the {\tt pexrav} code \cite{magdziarz95}, available in XSPEC, which calculates the reflected exponentially cutoff power law spectrum from neutral material. In all cases, we assumed  an inclination angle of 45$^{\rm o}$, $\Gamma_{\tt pexrav} =\Gamma_{\tt Nthcomp}$, and $E_{c, {\tt pexrav}}=kT_{e,\tt Nthcomp}$. The reflection amplitude, $R$, was negative so that {\tt pexrav} would output the reflection component only. We also considered neutral absorption with ${\rm N}_{\rm H}=10^{23}\,{\rm cm}^{-2}$ (similar to what has been measured in the past for the objects we studied - see Appendix), although it does not affect significantly the spectrum at the energies we consider. 

We downloaded the diagonal BAT survey instrument response from {HEASARC}\footnote{http://heasarc.nasa.gov/docs/swift/results/bs58mon}, and we used XSPEC to simulate {\it Swift}/BAT spectra for ${\Gamma}=1.3,1.7,2.1$. For each $\Gamma$, we considered three different electron temperatures: $kT_e=50,100,200$\,keV. For each pair of ($\Gamma, kT_e$) values, we used five different normalization values of {\tt Nthcomp}, from $A=1$ to $A=9$ photon cm$^{-2}$ s$^{-1}$, in steps of $\Delta A=2$ photon cm$^{-2}$ s$^{-1}$, to produce 5 simulated spectra (our results do not depend on the particular choice of the model normalization, as long as the ratio ($A_{\rm max}/A_{\rm min}$) is large enough to reproduce the observed maximum--to--minimum flux variations).  Finally, we added to the model a {\it constant} flux reflection component\footnote{This choice corresponds to the case of reflection from a distant, neutral reflector, i.e. the putative molecular torus. If the torus is a few pc away from the central source, the resulting reflection spectrum should be constant on time scales of a few years.}. The flux of the reflection component was such that $R_{\rm mean}=0.3, 1,$ and 1.5 in the case of the {\tt Nthcomp} ``mean" spectrum (i.e. the continuum when $A_{\rm mean}=5$ photon cm$^{-2}$ s$^{-1}$). Obviously, in this scenario, $R$ should decrease with increasing flux, since the flux of the reflection component remains constant.

\begin{figure*}
   \centering
   \includegraphics[bb=0 0 612 792,width=13.0cm,angle=270,clip]{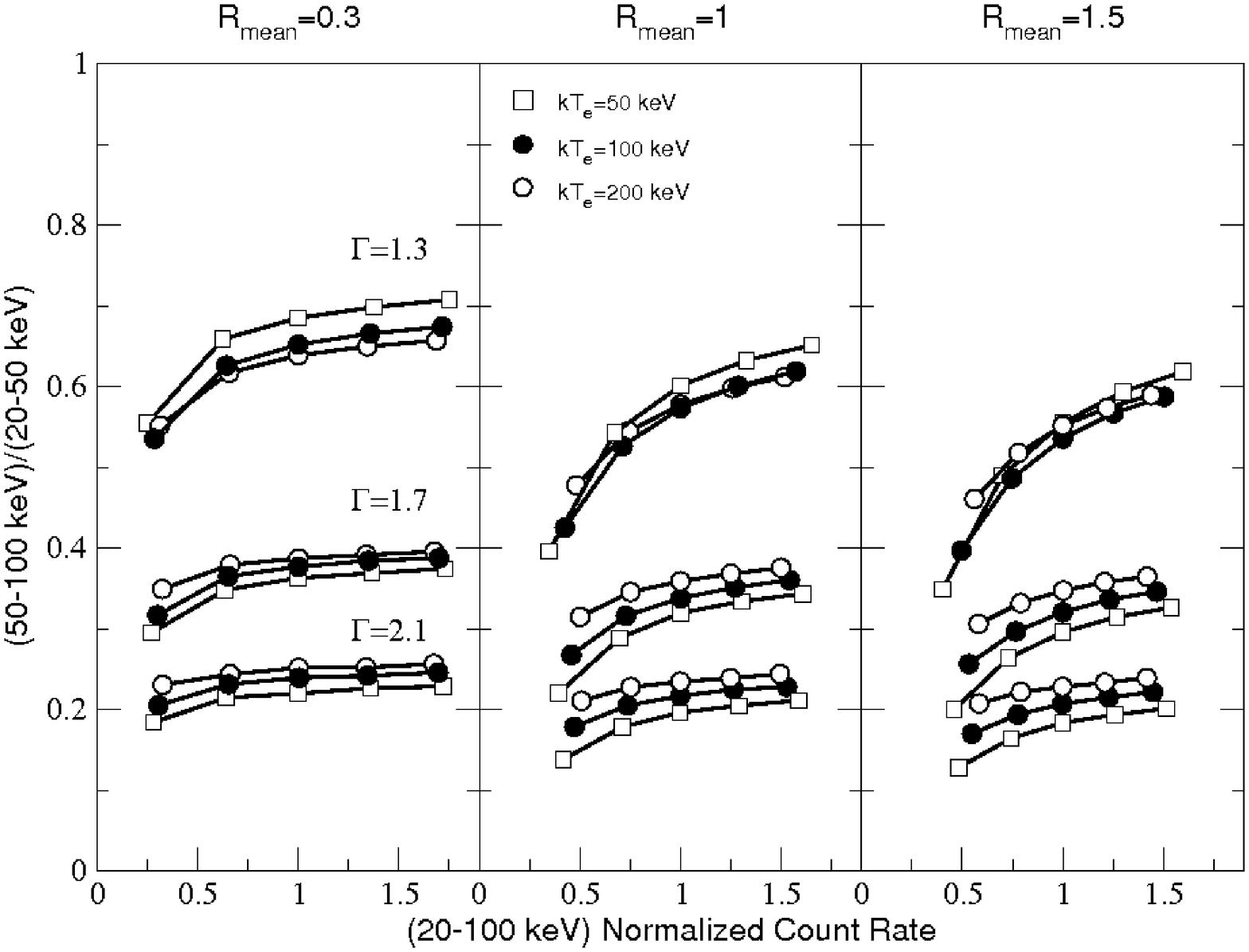}  
   \caption{Plot of model colour--flux diagrams in the case of a ``thermal Comptonisation plus a reflection component constant in flux" model. Different values for the (mean) reflection covering factor, power-law photon index and the temperature of the electrons in the corona have been considered: ${\rm R}_{\rm mean}=0.3,1,1.5$, from left to right, ${\Gamma}=1.3,1.7,2.1$, from top to bottom (in each panel), and ${\rm kT}_{\rm e}=50,100,200$\,keV, indicated with open squares, solid, and open circles, respectively. }          
              \label{bat_sim}%
    \end{figure*}

For each model spectrum we computed  the (20--100), (20--50) and (50--100) keV count rates, and the respective hardness ratio: (50--100 keV count rate)/(20--50 keV count rate). As a result, we were able to produce model colour--flux diagrams, which we plot in Fig. \ref{bat_sim}. The x--axis in these plots indicate the 20--100 keV model count rate normalized to the same band count rate of the ``mean" spectrum. The spread of the values along the x--axis is similar in the model and in the observed colour--flux diagrams (Fig.~\ref{colourflux}).  

The results for the $R_{\rm mean}=0.3, 1,$ and 1.5 case are plotted in the left, middle and right panel in this figure, respectively. Within each panel, the top, middle and bottom curves indicate the model colour--flux diagrams for $\Gamma=1.3, 1.7$ and $2.1$, respectively. Clearly, even the mean $HR$ value of a source can provide information about its $R_{\rm mean}$ and $\Gamma$. For example, the expected $HR$ value should be $\sim 0.2$ if $R_{\rm mean}=1.5$ and $\Gamma\sim 2.1$, as opposed to $\sim 0.55-0.7$ in the case of $R_{\rm mean}=0.3$ and $\Gamma\sim 1.3$. This is of course the case for the widest separated parameter values we considered, and the difference between the model colour--flux diagrams becomes less pronounced for smaller differences in the input physical parameters. Nevertheless, a comparison between the observed and model colour--flux diagrams plotted in  Figs.~\ref{colourflux} and Fig.~\ref{bat_sim} can provide interesting results. 

For example, the fact that $HR_{\rm NGC4151, NGC2110}\sim 0.4$ implies that $\Gamma<1.7$ and $R_{\rm mean}<1$ in both sources. If $\Gamma$ were steeper than 1.7, then $HR$ should be smaller than 0.4, while if $R_{\rm mean}\ge 1$, then we should had observed a significant positive correlation between $HR$ and flux (see the diagrams in Fig.~\ref{bat_sim} for $\Gamma>1.7$ and $R_{\rm mean}\ge 1$) which is not the case. 

Open squares, filled and open circles in each group of curves plotted in all panels of Fig.~\ref{bat_sim} correspond to $kT_e=50, 100$ and $200$ keV, respectively. Clearly, the model $HR$s do not depend significantly on $kT_e$. At a given flux, $HR$ decreases with decreasing $kT_e$, as the $50-100$ keV band flux decreases accordingly. These variations become more pronounced for $\Gamma>1.7$ and $R_{\rm mean}>1$ (see the two bottom group of curves in the middle and right panel of Fig.~5). In this case, when $kT_e$ changes from 200 keV to 50 keV, $HR$ decreases by $\sim 0.1$, which is comparable to the $HR$ variation when $\Gamma$ steepens from 1.7 to 2.1 (for the same  $kT_e$). 

In any case, the general trend of all the model diagrams in Fig. \ref{bat_sim} is that of a {\it positive} correlation between $HR$ and flux. Since the reflection component has a constant flux, and it is more pronounced at energies below 50 keV, the soft band count rate should remain roughly constant at low flux states; as the hard band flux will continue to decrease with decreasing flux, $HR$ decreases accordingly. And yet both the NGC~4151 and NGC~2110, where $HR$ remains constant, despite the significant flux variability of the sources.

The filled squares and triangles in the NGC~4151 colour--flux panel (Fig.~\ref{colourflux}) indicate the model colour--flux diagrams when $\Gamma=1.6$ and 1.7, respectively, $R_{\rm mean}=0.3$, and $kT_e$ decreases from 200 keV, in the lowest flux,  to 50 keV for the highest flux spectrum. The agreement between the observed and the model colour--flux diagrams is quite good. In the case of NGC~4151, any model diagram with a constant $kT_e$ will have a positive slope larger than the slope of the dashed and solid lines shown in this panel, and will obviously provide a worse ``fit" to the data. In other words, a $kT_e$ variation with flux is required in this source to reproduce its observed flat colour--flux diagram. In the presence of a constant flux reprocessed component, a power-law component which simply varies with flux does not result in a flat colour--flux diagram. The $kT_e$ variation between $50-200$ keV that our analysis implies is in excellent agreement with the results reported by Lubi{\'n}ski et~al. (2010) for the same source. A similar physical picture could also apply to with the observed colour--flux diagram of NGC~2110.

A positive correlation is opposite to what we observe in NGC~4388 (and NGC~4945 and IC~4329 as well). Within the context of the model we considered, we can reproduce an $HR-$flux anti-correlation only if the intrinsic $\Gamma$ correlates {\it positively} with flux. Filled circles, connected with a solid line,  in the observed colour--flux diagram of NGC~4388 in Fig.~\ref{colourflux} indicate the model $HR$ values when  $R_{\rm mean}=0.3$, $kT_e=100$ keV, and $\Gamma$ increases from 1.3, in the lowest flux, to 2.1, for the highest flux model spectrum. Although the model colour-flux diagram slope is steeper than the observed best-fit slope (a result which implies that the intrinsic $\Delta\Gamma$ in NGC~4388 is probably less than the $\Delta\Gamma_{\rm model}=0.8$ we considered), it is obvious that the observed spectral variations in NGC~4388 (and to a lesser extent in NGC~4945 and IC~4329 as well) can be explained by an {\it intrinsic} spectral slope steepening with increasing flux. 

\section{Discussion and conclusions}  \label{discuss}

We used the 58-month long {\it Swift}/{\rm BAT} monitoring light curves to study the flux and spectral variability of the five brightest Seyfert galaxies in the catalogue of Baumgartner et al. (2011). Our main results are summarized below.

\begin{itemize}

\item{Both the 20--50 and 50--100 keV band light curves, of all five objects, are significantly variable on time scales as short as 1--2 days. Their average variability amplitude is of the order of $\sim  30-50$\% of the mean source flux.  We also found evidence that the variability amplitude scales inversely proportional with the BH mass of the objects, a trend which has also been observed in AGN, but at energies below 10 keV.}

\end{itemize}

Since the column of the neutral material which obscures the X--ray source in these objects is less than $10^{24}$ cm$^{-2}$, it cannot affect significantly the X-ray emission of the sources above 20 keV.  Consequently, the {\it Swift}/BAT light curves can reveal the ``true" variability behaviour of the central source in these, without the complications associated with the effects of the absorbing material at energies below $\sim 10$ keV. Therefore, the observed variability must be intrinsic to the central source. Furthermore, the similarity of the variability amplitudes at energies above 20 keV and below 10 keV in AGN (for a given BH mass) implies that the main driver of the observed flux variations below $\sim$ 20 keV on time scales longer than 1--2 days may also be the intrinsic normalisation variations of the source continuum. 

\begin{itemize}

\item{The soft (20--50 keV) and hard band (50--100 keV) light curves, in all objects, are well correlated. We do not detect any delays down to 1 day, which is the smallest time scale we can sample with the light curves we used. In other words, similar variations appear, almost simultaneously, in both energy bands.} 

\end{itemize}

In the case of NGC~4151, which is the object with the highest signal-to-noise ratio light curves, we detected a soft band delay, at time scales longer than $\sim 10$ days. This is the first time that such a delay within the X--ray band has been observed in an AGN, but it is not easy to quantify the significance of this result, and it is not clear what are its implications. If real, such a result implies that a physical mechanism (perhaps a change in the temperature of the hot plasma in the corona?) affects first the high energy part of the spectrum of the source, and then propagates to softer energies.

\begin{itemize}

\item{A hardness ratio analysis suggests that, despite the significant flux variations, the shape of the 20--100 keV spectrum remains constant in NGC~4151 and NGC~2110. On the other hand, we detected significant spectral variations in NGC~4388: the hardness ratios decrease as the flux increases. A similar trend is also observed in NGC~4945 and IC~4329, but its significance is marginal.}

\end{itemize}

There is not a clear understanding of the physical parameters that drive the spectral evolution of  AGN. As reported in the Appendix, previous studies of the sources in the sample have revealed hard X-ray spectral variations due to changes of either the photon index  of the X--ray continuum, or of the reflection amplitude, and/or of the temperature of the Comptonizing electrons. 

The use of  the (20--50) and (50--100) keV band light curves to compute the $HR$s means that we are not very sensitive in the detection of $kT_e$ variations, as long as $kT_{e}> 50$ keV. Interestingly though, the flatness of the colour--flux diagrams of NGC~4151 and NGC~2110 suggests that the electron temperature should decrease with increasing flux in these objects. This in agreement with the results of Lubi{\'n}ski et~al. (2010) in the case of NGC~4151. Furthermore, the average $HR$ value in these objects suggest that $R_{\rm mean}$ should be less than unity, and $\Gamma_{\rm intrinsic}$ less than $\sim 1.7$. 

Comparison of the colour-flux diagrams of NGC~4388  (and perhaps in NGC~4945 and IC~4329 as well) with the predictions of thermal Comptonization models supports the view that  the main driver of spectral variability in this source is intrinsic spectral slope variations: $\Gamma_{\rm intrinsic}$ correlates positively with the source flux, while the reflection component flux remains constant. As a result, $R_{\rm obs}$ should also decrease with increasing flux.

It is difficult to identify the physical parameter responsible for the $\Gamma_{\rm intrinsic}-$flux correlation that we observe in NGC~4388. An anti-correlation between $kT_{e}$ and source flux, like the one observed in NGC~4151, could explain it, if the optical depth of the corona remains constant: in this case, a decrease in the temperature of the corona results in a steeper continuum. . On the other hand, the same NGC~4151 observations suggest that, as $kT_e$ varies, the optical depth of the corona also varies so that the Compton parameter $y$ (and hence $\Gamma$ as well) remains constant (for this source). Therefore, flux related optical depth variations could also explain the $\Gamma-$ flux correlation, if $kT_e$ remains constant. However, due to the fact that previous studies have indicated $kT_e$ variations in NGC~4388  (see Appendix), we believe that variations of the coronal temperature are the main driver of the spectral variability in this object. 

A ``softer when brighter" behaviour is commonly detected in Seyferts at energies below 10 keV, both on short and long time scales. Sobolewska \& Papadakis (2009) have found that the observed spectral variations of X--ray bright AGN in the 2--20 keV band can be explained if $\Gamma_{\rm intrinsic}\propto F_{\rm 2-10 keV}^{0.1}$. If that were the case, the max-to-min flux ratio of $\sim 7$ in the case of NGC~4388 should imply a $\Gamma_{\rm max}/\Gamma_{\rm min}$ ratio of $\sim 1.2$. The dashed line in the bottom panel of Fig.~\ref{colourflux} indicates the expected colour--flux diagram when $\Gamma=1.3$, for the lowest flux, and $\Gamma=1.7$ for the highest flux spectrum (so that $\Gamma_{\rm max}/\Gamma_{\rm min}\sim 1.3$). The model colour--flux diagram has an amplitude which is higher than the amplitude of the observed diagram, most probably because $R_{\rm mean}>0.3$ in this source. However, the slope of both the model and the observed colour--flux diagram are quite similar. Therefore, our results, are in agreement and support those of Sobolewska \& Papadakis (2009). 

The absence of significant spectral variations in NGC~4151 and NGC~2110, although rare among Seyfers at lower energies, is not unique. For example, Sobolewska \& Papadakis (2009) noticed that ``NGC 5548 displayed limited spectral variations for its flux variability", and Papadakis, Reig \& Nandra (2003) observed the same behaviour in PG 0804+761. One important difference between NGC~4151 and NGC~2110 and the other three sources in the sample is that they accrete at $\sim 1\%$ (or less) of the Eddington limit (see Appendix), as opposed to $\sim 10\%$ (or higher) for the other sources. Recently, Sobolewska et al. (2011) studied the long term ``spectral slope--flux" evolution of two well studied GBHs, namely GRO J1655-40 and GX 339-4. Their results indicate that at accretion rates  $\sim 0.01$ (or smaller) of the Eddington limit, the X--ray spectral slope is $\sim 1.3-1.6$, and it remains roughly constant over a large flux range (see the middle panel of their Fig.~2; in particular the panel with the GRO J1655-40 data). This is exactly the case with NGC~4151 and NGC~2110 as well: they both have an accretion rate lower than $\sim 1$\%, a spectral slope flatter than $\sim 1.7$ (see e.g. Lubi{\'n}ski et~al. 2010, Winter et~al. 2009) and do not exhibit significant spectral variations. These similarities raise the issue of different ``spectral states" in AGN, just like in GBHs, with NGC~4151 and NGC~2110 being ``hard-state" systems. 
 
\begin{acknowledgements}

This research has made use of data obtained from the High Energy Astrophysics Science Archive Research Center (HEASARC), provided by NASA's Goddard Space Flight Center. MCG, IEP and FN acknowledge support by the EU FP7-REGPOT 206469 grant. 

\end{acknowledgements}

\section{APPENDIX} 

We present below results from previous studies of the hard X--ray spectrum of the sources in the sample. 

\subsection{NGC~4151}

NGC~4151 is a nearby Seyfert 1.5 galaxy, and the brightest persistent AGN in the 20-100\,keV energy band (after the radio-loud blazar Cen A).  It hosts a BH with a mass of $4.6^{+0.6}_{-0.5}{\times}10^{7}\,{\rm M}_{\odot}$, as estimated from reverberation mapping studies \citep{bentz06},
and radiates at a rate of of ${\rm L}_{\rm Bol}/{\rm L}_{\rm EDD}=0.014$ \citep{crenshaw07}. 

Its spectrum has been extensively studied from radio wavelengths to hard X-rays. Its energy spectrum above 20 keV is well fitted by a power-law model (PL, hereafter), with an exponential cut-off above energies 50--200 keV (see e.g. Piro et~al., 1999 and 2000; Beckmann et~al., 2005 and 2009; Lubi{\'n}ski et~al.,2010). Lubi{\'n}ski et~al. (2010) in particular, have suggested that the cut-off energy, $E_{\rm c}$, anti-correlates with the source flux, increasing from $\sim 5-80$ keV, at high fluxes, to $\sim 100-200$ keV, when the source is at ``dim flux state".  The spectral slope of the PL component, $\Gamma$, is rather hard (compared to other, X--ray bright, Seyfert galaxies), with best-fit values being $\sim 1.4-1.6$ (see e.g. Zdziarski et~al., 1996; Petrucci et~al., 2001; Winter et~al., 2009). de Rosa et~al. (2007) have suggested that there may exist flux related, intrinsic spectral slope variations in this source. In addition to the PL component,  {\it Ginga} observations first established the presence of a ``reflection hump" at energies above 10 keV (Zdziarski et~al., 1996). This component has been detected by all major hard X--ray satellites since then, but with a variable amplitude, $R$, which ranges between 0.01 and almost 1 (i.e. the value expected in the case of a point source located on top of the accretion disc at relatively large height). Finally, it is well established that the X-ray spectrum of NGC~4151 is also affected by the presence of neutral material, perhaps with a variable covering factor, and a column density, N$_{\rm H}$, less than $\sim 10^{23}$ cm$^{-2}$ (see e.g. Lubi{\'n}ski et~al., 2010, and references therein).

\subsection{NGC~4945}

NGC~4945 is a nearby (3.7\,Mpc; Mauersberger et~al.1996) Seyfert 2 galaxy, with a BH mass estimate of ${\approx}1.4{\times}10^{6}\,{\rm M}_{\odot}$ \citep{greenhill97}, which radiates at 
a rate of ${\rm L}_{\rm Bol}/{\rm L}_{\rm EDD}=0.10-0.60$ \citep{madejski00}. {\it Ginga} observations revealed the presence of a hard X--ray source in this galaxy, which is heavily absorbed at energies below 10 keV by neutral material with an N$_{\rm H}$ as high as $\sim 10^{24}$ cm$^{-2}$ (Iwasawa et al., 1993).  Large amplitude X-ray variations at energies above 10 keV, on time-scales less than ${\le}1-2$\,days (and as short as $\sim 10^4$ s) have been detected by {\it RXTE}, {\it BeppoSAX} and {\it Suzaku} observations \citep {madejski00,guainazzi00,takeshi08}. These flux variations do not appear to be correlated with spectral variability as well (Guainazzi et al., 2000). Best-fit $\Gamma$ values range between 1.4 and 1.8 (see e.g. Iwasawa et~al, 1993; Done et~al, 1996; Madejski et al., 2000; Guainazzi et al., 2000; Beckmann et~al., 2009; Winter et~al., 2009). In addition, the high energy cut-off in the X--ray spectrum is constrained to be in the range $\sim 100-300$ keV, while the reflection component appears to have a rather small value of $R\sim 0.06$ (Done et~al, 2003). 

\subsection{NGC~2110}

NGC~2110 is a nearby ($z=0.007579$) Seyfert 2 galaxy. It hosts a black hole with a mass of $2{\times}10^{8}\,{\rm M}_{\odot}$ \citep{woo02} and radiates at ${\rm L}_{\rm Bol}/{\rm L}_{\rm EDD}=10^{-3}-10^{-2}$ \citep{evans07}. {\it BeppoSAX} observations \citep{malaguti99} revealed a steep power-law spectrum of ${\Gamma}=1.9$, and the presence of a neutral absorber with ${\rm N}_{\rm H}{\approx}7{\times}10^{23}\,{\rm cm}^{-2}$, partially covering the source. The  same observations implied a rather low value for the reflection covering factor (${\rm R}{\le}0.17$ or $\le 0.5$,  if ${\rm E}_{\rm c}=1000$ or $50$\,keV, respectively). The {\it Second INTEGRAL catalogue} \citep{beckmann09} yielded the best-fit values of  ${\Gamma}=2$ and ${\rm N}_{\rm H}=4.3{\times}10^{22}\,{\rm cm}^{-2}$ for the PL slope and the column density of the absorbing neutral material, intrinsic to the source. Spectral studies of the {\it Swift}/{\rm BAT} spectra \citep{winter09} provided the following estimates for the photon index and column density of the neutral absorber: ${\Gamma}=1.54^{+0.08}_{-0.07}$ and ${\rm N}_{\rm H}=2.84^{+0.19}_{-0.16}{\times}10^{22}\,{\rm cm}^{-2}$.

\subsection{IC~4329}

IC~4329 is a relatively nearby (z=0.0157; Wilson \& Penston, 1979), X--ray luminous AGN. It hosts a black hole with a mass of $1.3^{+1.0}_{-0.3}{\times}10^{8}\,{\rm M}_{\odot}$ and radiates at a ratio of ${\rm L}_{\rm Bol}/{\rm L}_{\rm EDD}=0.21^{+0.06}_{-0.10}$ \citep{markowitz09}. 
 {\it BeppoSAX} observations have shown significant variations in the 0.1--100\,keV flux emission from the source (Perola et al., 1999), which were not associated with significant spectral variations. 
 {\it COMPTON/OSSE} observations have indicated the presence of a high energy cut-off in the hard X--ray spectrum of the source, at energies $250\,{\rm keV}{\le}{\rm E}_{\rm c}{\le}1700\,{\rm keV}$ \citep{madejski95}. Subsequent {\it INTEGRAL} observations have indicated a smaller cut-off energy of $\sim 80$ keV (Beckmann et al., 2009). The best-fit X--ray spectral slope values at hard energies range from $\sim 1.4$ to $\sim 1.8$ (see e.g. Beckman et al., 2009; Winter et al., 2009). There are indications of neutral absorption intrinsic to the source, with an N$_{\rm H}<10^{24}$ cm$^{-2}$, and of the presence of a reflection hump with a strength of $R\sim 0.4-1.2$ (see e.g. Miyoshi et al., 1998; Done \& Madejski, 2000). 

\subsection{NGC~4388}

NGC~4388 is a nearby ($z=0.00842$) nearly edge-on ($i{\simeq}78^{\circ}$) spiral galaxy hosting a Seyfert 2 nucleus \citep{phillips82,filippenko85}. 
It hosts a black hole with a mass of $6{\times}10^{6}\,{\rm M}_{\odot}$ \citep{woo02} and radiates at ${\rm L}_{\rm Bol}/{\rm L}_{\rm EDD}=0.1$ \citep{elvis04}.
{\it SIGMA} \citep{paul91}, {\it OSSE} \citep{johnson93} and {\it BeppoSAX} \citep{butler91,boella97} observations during 13 years showed no spectral
shape variations of the PL. Subsequent {\it INTEGRAL} \citep{beckmann04} and {\it Suzaku} \citep{shirai08} observations revealed that the flux increased by 
a factor of 1.4 and 1.5, respectively (with a month to half-day timescales, respectively). The best-fit X-ray spectral slope values at hard energies range from 
${\approx}1.3$ to ${\approx}1.8$ \cite{beckmann09,winter09}, with column density of ${\rm N}_{\rm H}=2-3{\times}10^{23}\,{\rm cm}^{-2}$. The hard X-ray spectrum does not show 
a strong cut-off in the X-rays at $E{\le}100$\,keV, with the best estimate being ${\rm E}_{\rm c}=95^{+26}_{-17}$\,keV, from {\it INTEGRAL} observations 
\cite{beckmann09}. The reflection component was detected during the {\it Suzaku} observations and the reflection covering factors was 
constrained to be ${\rm R}=1.3-1.5$. There is barely a spectral shape variation between the high and the low-flux states, with the most important change being the 
normalization of the underlying power-law continuum from the nucleus.

\end{document}